\documentclass[journal,onecolumn]{IEEEtran}

\usepackage{amsmath,amstext,amsfonts,amssymb,amsthm,bbold} 
\usepackage{graphicx,subfigure,color} 
\usepackage{cite,hyperref}
\usepackage{booktabs,multirow} 

\def \A {\mathbf{A}}

\def \B {\mathbf{B}}
\def \b {\mathbf{b}}
\def \C {\mathbf{C}}

\def \e {\mathbf{e}}

\def \G {\mathbf{G}}
\def \g {\mathbf{g}}
\def \h {\mathbf{h}}
\def \H {\mathbf{H}}
\def \I {\mathbf{I}}

\def \R {\mathbf{R}}

\def \S {\mathbf{S}}

\def \u {\mathbf{u}}
\def \v {\mathbf{v}}

\def \W {\mathbf{W}}
\def \w {\mathbf{w}}
\def \x {\mathbf{x}}
\def \X {\mathbf{X}}
\def \Y {\mathbf{Y}}
\def \y {\mathbf{y}}
\def \Z {\mathbf{Z}}
\def \z {\mathbf{z}}

\def \Ccal {\mathcal{C}}

\def \Hcal {\mathcal{H}}

\def \Ncal {\mathcal{N}}
\def \Ocal {\mathcal{O}}

\def \Vcal {\mathcal{V}}

\def \Cbb {\mathbb{C}}
\def \Ebb {\mathbb{E}}

\def \Pbb {\mathbb{P}}
\def \Rbb {\mathbb{R}}

\def \Zbb {\mathbb{Z}}

\def \drm {\mathrm{d}}
\def \erm {\mathrm{e}}
\def \irm {\mathrm{i}}

\def \chibs {\boldsymbol{\chi}}
\def \deltabs {\boldsymbol{\delta}}
\def \epsilonbs {\boldsymbol{\epsilon}}

\def \phibs {\boldsymbol{\phi}}

\def \xibs {\boldsymbol{\xi}}

\def \Phibs {\boldsymbol{\Phi}}
\def \Sigmabs {\boldsymbol{\Sigma}}

\def \Xibs {\boldsymbol{\Xi}}

\def \det {\mathrm{det}}

\def \Tr {\mathrm{Tr}\,}
\def \tr {\mathrm{Tr}\,}

\DeclareMathOperator*{\argmax}{argmax}

\DeclareMathOperator*{\dg}{dg}
\renewcommand{\Im}{\mathrm{Im}}

\renewcommand{\exp}{\mathrm{exp}}

\newtheorem{assumption}{Assumption}
\newtheorem{corollary}{Corollary}
\newtheorem{theorem}{Theorem}

\newtheorem{lemma}{Lemma}
\newtheorem{remark}{Remark}

\newtheorem{proposition}{Proposition}

\begin{document}

\title{On the detection of low-rank signal in the presence of spatially uncorrelated noise: a frequency domain approach.}

\author
{
  A. Rosuel, \IEEEmembership{Student Member, IEEE}, 
  P. Vallet, \IEEEmembership{Member, IEEE},
  P. Loubaton, \IEEEmembership{Fellow, IEEE},
  and
  X. Mestre, \IEEEmembership{Senior Member, IEEE}
  \thanks
  {
    A. Rosuel and P. Loubaton are with Laboratoire d'Informatique Gaspard Monge (CNRS, Univ. Gustave-Eiffel),
    5 Bd. Descartes 77454 Marne-la-Vall{\'e}e (France),
    \{alexis.rosuel, philippe.loubaton\}@univ-eiffel.fr
  }
  \thanks
  {
    P. Vallet is with Laboratoire de l'Int{\'e}gration du Mat{\'e}riau au Syst{\`e}me (CNRS, Univ. Bordeaux, Bordeaux INP),
    351, Cours de la Lib{\'e}ration 33405 Talence (France),
    pascal.vallet@bordeaux-inp.fr	
  }
  \thanks
  {
    X. Mestre is with Centre Tecnol\`{o}gic de Telecomunicacions de Catalunya (CTTC),
    Av. Carl Friedrich Gauss 08860 Castelldefels, Barcelona (Spain), 
    xavier.mestre@cttc.cat
  }
  \thanks
  {
    This work was partially supported by project ANR-17-CE40-0003.
    The material of this paper was partly presented in the conference paper \cite{Rosuel2020}
  }
}

\maketitle

\begin{abstract}
  This paper analyzes the detection of a $M$--dimensional useful signal modeled as the output of a $M \times K$ MIMO filter driven by a $K$--dimensional
  white Gaussian noise, and corrupted by a $M$--dimensional Gaussian noise with mutually uncorrelated components.
  The study is focused on frequency domain test statistics based on the eigenvalues of an estimate of the spectral coherence matrix (SCM), obtained as a
  renormalization of the frequency-smoothed periodogram of the observed signal. If $N$ denotes the sample size and $B$ the smoothing span,
  it is proved that in the high-dimensional regime where $M,B,N$ converge to infinity while $K$ remains fixed, the SCM behaves
  as a certain correlated Wishart matrix. Exploiting well-known results on the behaviour of the eigenvalues of such matrices,
  it is deduced that the standard tests based on linear spectral statistics of the SCM fail to detect the presence of
  the useful signal in the high-dimensional regime. A new test based on the SCM, which is proved to be consistent, is also proposed, and its
  statistical performance is evaluated through numerical simulations.
\end{abstract}


\section{Introduction}
\label{section:introduction}

\IEEEPARstart{D}{etecting} the presence of an unknown multivariate signal corrupted by noise is one of the fundamental problems in signal processing,
which is found in many applications including array and radar processing, wireless communications, radio-astronomy or seismology among others.
In a statistical framework, this problem is usually formulated as the following binary hypothesis test, where the objective is to discriminate between the null hypothesis $\Hcal_0$ and the alternative hypothesis $\Hcal_1$ defined as
\begin{align}
  \label{eq:det_test}
  \Hcal_0&: (\y_n)_{n \in \Zbb} = (\v_n)_{n \in \Zbb}
           \notag\\
  \Hcal_1&: (\y_n)_{n \in \Zbb} = (\u_n)_{n \in \Zbb} + (\v_n)_{n \in \Zbb}
\end{align}
where $(\y_n)_{n \in \Zbb}$ is the $M$-variate observed signal, and where $(\u_n)_{n \in \Zbb}$ and $(\v_n)_{n \in \Zbb}$ represent a non observable signal of interest
and the noise respectively, both modeled in this paper as mutually independent zero-mean complex Gaussian stationary time series. 

Without further knowledge on the covariance function of $(\v_n)_{n \in \Zbb}$ and/or $(\u_n)_{n \in \Zbb}$, or access to ``noise only'' samples, the test problem \eqref{eq:det_test} is ill-posed, even for temporally white time series $(\v_n)_{n \in \Zbb}$ and $(\u_n)_{n \in \Zbb}$, and one needs to exploit additional information on the covariance structure of the useful signal and noise. One common assumption, widely used in the context of array processing and multi-antenna communications, is to consider that the noise $(\v_n)_{n \in \Zbb}$ is spatially uncorrelated. Moreover, when the 
receive antennas are not calibrated, it is reasonable to assume that the spectral densities 
of the components of the noise may not coincide, see e.g. \cite{Ramirez2011}, \cite{sala2016multiantenna}, \cite{Boonstra2003}, \cite{Leshem2001}. This will be the context of the present paper.

A first class of tests is based on the observation that the noise is spatially uncorrelated if and only 
if the matrices $\R_{\v}(\ell) = \Ebb[\v_n\v_{n-\ell}^*]$ are diagonal for all $\ell \in \Zbb$, whereas if the useful signal $(\u_n)_{n \in \Zbb}$ is assumed spatially correlated, 
$\R_{\u}(\ell) = \Ebb[\u_n\u_{n-\ell}^*]$ is non-diagonal for some $\ell \in \Zbb$.
Under this assumption, the problem in \eqref{eq:det_test} can be formulated as the following correlation test:
\begin{align}
  \label{eq:cor_test}
  \Hcal_0&: \R_{\y}(\ell) = \dg\left(\R_{\y}(\ell)\right) \text{ for all } \ell \in \Zbb
           \notag\\
  \Hcal_1&: \R_{\y}(\ell) \neq \dg\left(\R_{\y}(\ell)\right) \text{ for some } \ell \in \Zbb
\end{align}
where $\R_{\y}(\ell) = \Ebb[\y_n\y_{n-\ell}^*]$ and $\dg\left(\R_{\y}(\ell)\right) = \R_{\y}(\ell) \odot \I_M$, where $\odot$ is the element-wise (Hadamard) product and $\I_M$ the $M \times M$ identity matrix.
A number of previous works developed lag domains tests that specifically tackle the above problem, see e.g. \cite{haugh1976checking}, \cite{hong1996testing}, \cite{li1994robust}, \cite{elhimdiduchesneroy2003}, 
\cite{Ramirez2010}, \cite{Klausner2014}. Also relevant are the approaches in \cite{Ramirez2011} and \cite{sala2016multiantenna}, where the possible useful signal is supposed to be the output of a filter driven by a low-dimensional white noise sequence.

Our focus here is on another type of formulation, referred to as \emph{frequency domain approach}, which consists in rewriting problem \eqref{eq:cor_test} as
\begin{align}
  \label{eq:spec_test}
  \Hcal_0&: \S_{\y}(\nu) = \dg\left(\S_{\y}(\nu)\right) \text{ for all } \nu \in [0,1]
           \notag\\
  \Hcal_1&: \S_{\y}(\nu) \neq \dg\left(\S_{\y}(\nu)\right) \text{ for some } \nu \in [0,1]
\end{align}
where $\S_{\y}(\nu)$ is the $M \times M$ spectral density matrix of $(\y_n)_{n \in \Zbb}$ at frequency $\nu$, defined by
\[
    \S_\y(\nu) = \sum_{k\in\Zbb}\R_\y(k)e^{-2\irm\pi\nu k}.
\]
This problem is equivalent to testing whether the \emph{spectral coherence matrix} (see for instance \cite[Chapter 7-6]{Brillinger1981}, \cite[Chapter 5.5]{Koopmans1995})
\begin{align}
\label{equation:definition_C_y}
  \C_{\y}(\nu) = \dg\left(\S_{\y}(\nu)\right)^{-\frac{1}{2}} \S_{\y}(\nu) \dg\left(\S_{\y}(\nu)\right)^{-\frac{1}{2}}
\end{align}
is equal to $\I_M$ for all frequencies $\nu \in [0,1]$.
In this approach, usual test statistics are mostly based on consistent sample estimates of $\S_{\y}(\nu)$ or $\C_{\y}(\nu)$ that are compared to a diagonal matrix or to the identity $\I_M$ respectively. Previous works that developed this approach include  \cite{wahba1971some}, \cite{taniguchi1996nonparametric}, \cite{eichler2007frequency}, 
\cite{eichler2008testing}. In particular, \cite{wahba1971some} considered the following frequency smoothed-periodogram estimator $\hat{\S}_{\y}(\nu)$ defined by 
\begin{align}
\label{equation:definition_S_y}
  \hat{\S}_{\y}(\nu) =
  \frac{1}{B+1} \sum_{b=-\frac{B}{2}}^{\frac{B}{2}} \xibs_{\y}\left(\nu + \frac{b}{N}\right) \xibs_{\y}\left(\nu + \frac{b}{N}\right)^*
\end{align}
with  $\xibs_{\y}(\nu) = \frac{1}{\sqrt{N}} \sum_{n=0}^{N-1} \y_n e^{-2 i \pi n \nu}$ the renormalized finite Fourier transform of $(\y_n)_{n=0,\ldots,N-1}$, $B$ the \emph{smoothing span}, assumed to be an even number,  and where $\xibs_{\y}\left(\nu + \frac{b}{N}\right)^*$ is the conjugate transpose of the vector $\xibs_{\y}\left(\nu + \frac{b}{N}\right)$. \cite{wahba1971some} was devoted to the study of the limiting distribution of
\[
    \log\left\{\prod_{i=1}^P \det(\hat{\S}_\y(\nu_i))/\prod_{m=1}^M \hat{s}_{m,m}(\nu_i)\right\}
\]
for some properly defined subset of frequencies $(\nu_i)_{i=1,\ldots,P}$, where $\hat{s}_{m,m}(\nu) = \left(\hat{\S}_{\y}(\nu)\right)_{m,m}$.  When $M=2$, \cite{eichler2007frequency} considered a general kernel estimator of $\S_\y(\nu)$:
\[
    \tilde{\S}_{\y}(\nu) = \sum_{b=-\frac{N}{2}}^{\frac{N}{2}} w_N\left(\frac{b}{N}\right) \xibs_{\y}\left(\nu + \frac{b}{N}\right) \xibs_{\y}\left(\nu + \frac{b}{N}\right)^*
\]
where $w_N$ is a weight function satisfying some specific properties, and a test statistic of the form 
\[
    \frac{1}{N}\sum_{n=1}^N\frac{|(\tilde{\S}_{\y})_{12}(\nu)|^2}{(\tilde{\S}_{\y})_{11}(\nu)(\tilde{\S}_{\y})_{22}(\nu)}
\]
which is proven to be, after proper recentring and renormalization, asymptotically normally distributed. Finally, \cite{taniguchi1996nonparametric} and \cite{eichler2008testing} considered the more general class of test statistics, defined by:
\[
    \int_{-1/2}^{1/2} K\left((\tilde{\S}_{\y})_{12}(\nu)\right) d\nu  \text{ and } \int_{-1/2}^{1/2} \left\|\psi\left((\tilde{\S}_{\y})_{12}(\nu),\nu\right)\right\|^2 d\nu
\]
for some well-defined functions $K$ and $\psi$, and where $\|\cdot\|$ is the Euclidian norm. They proved that these quantities asymptotically follow normal distributions. In the present paper, we focus on the natural estimator (see e.g. \cite[Chapter 7-6]{Brillinger1981}, \cite[Chapter 8-4]{Koopmans1995}) of $\C_{\y}$, defined by
\begin{align}
\label{equation:definition_hat_C_y}
  \hat{\C}_{\y}(\nu) =
  \dg\left(\hat{\S}_{\y}(\nu)\right)^{-\frac{1}{2}} \hat{\S}_{\y}(\nu) \dg\left(\hat{\S}_{\y}(\nu)\right)^{-\frac{1}{2}}
\end{align}
where $\hat{\S}_{\y}(\nu)$ is the frequency-smoothed periodogram estimate defined by (\ref{equation:definition_S_y}). 
Note that adding a weight to the matrices $\xibs_{\y}(\nu + \frac{b}{N}) \xibs_{\y}(\nu + \frac{b}{N})^*$ leads to a more general class of estimators of $\S_\y(\nu)$. The study of this more general class of estimators involves different techniques and random matrix models than the ones used here, and is therefore out of the scope of this paper.

\subsection{Low vs High-dimensional regime}
The performance of the test statistics developed in the above mentioned previous works 
is usually studied in the \emph{low-dimensional regime} where $N \to \infty$ and $M$ is fixed. It is well known (see for instance \cite{Brillinger1981}) that $ \hat{\S}_{\y}(\nu)$ and $ \hat{\C}_{\y}(\nu)$ are consistent estimates if $B \rightarrow +\infty$ and  $\frac{B}{N} \to 0$. Under mild assumptions on the memory of the time series $(\y_n)_{n \in \Zbb}$, $\hat{\C}_{\y}(\nu)$ is a consistent and asymptotically normal estimate of $\C_{\y}(\nu)$, which can in turn be used to study the asymptotic performance of the various tests based on $ \hat{\C}_{\y}(\nu)$. In practice, the above asymptotic 
regime allows to predict the actual performance of the tests quite accurately, provided the ratio $\frac{M}{N}$ is small enough. If this condition is not met, test statistics based on $\hat{\C}_{\y}(\nu)$ may be of delicate use, as the choice of the smoothing span $B$ must meet the constraints $\frac{B}{M}$ much larger than 1 (because $B$ is supposed to converge towards $+\infty$) as well as $\frac{B}{N}$ small enough (because $\frac{B}{N}$ is supposed to converge towards $0$). 

Nowadays, in many practical applications involving high-dimensional signals and/or a moderate sample size, the ratio $\frac{M}{N}$ may not be small enough to be able to choose $B$ so as to meet $\frac{B}{M}$ much larger than 1 and  $\frac{B}{N}$ small enough. Therefore, the results obtained in the low-dimensional regime may fail to provide accurate predictions of the behaviour of the aforementioned test statistics. In this situation, one may rely on the more relevant \emph{high-dimensional regime} in which $M,B,N$ converge to infinity such that
$\frac{M}{B}$ converges to a positive constant while $\frac{B}{N}$ converges to zero.

In comparison to the low-dimensional regime, the literature concerning correlation tests for the frequency domain in the high-dimensional regime is quite scarce.
Recent results obtained in \cite{loubaton2020large} show that under hypothesis $\Hcal_0$, the empirical eigenvalue distribution of the spectral coherence estimate $\hat{\C}(\nu)$ behaves in the high-dimensional regime as the well-known Marcenko-Pastur distribution \cite{Marcenko1967}.
The result of \cite{loubaton2020large} allows to predict the performance under $\Hcal_0$ of a large class of test statistics based on
\begin{align}
  L_{\varphi}(\nu) = \frac{1}{M} \sum_{m=1}^M \varphi\left(\lambda_m(\hat{\C}_{\y}(\nu))\right)
  \notag
\end{align}
where $\lambda_1(\hat{\C}_{\y}(\nu)),\ldots,\lambda_M(\hat{\C}_{\y}(\nu))$ are the eigenvalues of $\hat{\C}_{\y}(\nu)$, and $\varphi$ belongs to a certain functional class.
Such family of statistics $L_{\varphi}$, called \emph{linear spectral statistics} (LSS) of $\hat{\C}_{\y}(\nu)$, include in particular the choice $\varphi(x) = \log x$, i.e. $L_{\varphi}(\nu) = \frac{1}{M} \log \mathrm{det}\hat{\C}_{\y}(\nu)$ and the choice  $\varphi(x) = (x-1)^2$, i.e. 
 $L_{\varphi}(\nu) = \frac{1}{M} \| \hat{\C}_{\y}(\nu) - \mathbf{I}_M \|^{2}_{F}$, where $\|\cdot\|_F$ represents the Frobenius norm.

In this paper, we consider the study of the eigenvalues of $\hat{\C}_{\y}(\nu)$ in the high-dimensional regime under the special alternative $\Hcal_1$ for which the useful signal $(\u_n)_{n\in \Zbb}$ is modeled as the output of a stable MIMO filter driven by a $K$--dimensional white complex Gaussian noise. In the context where the intrinsic dimension $K$ is fixed while $M,N,B \to \infty$, it is shown that the empirical eigenvalue distribution of
$\hat{\C}_{\y}(\nu)$ still converges to the Marcenko-Pastur distribution, showing that the test statistic based on $L_{\varphi}(\nu)$ is unable to discriminate between hypotheses $\Hcal_0$ and $\Hcal_1$ in the high-dimensional regime. Nevertheless, we also prove that, provided that the signal-to-noise ratio is large enough, the  largest eigenvalue of $\hat{\C}_{\y}(\nu)$
asymptotically splits from the support of the Marcenko-Pastur distribution. We can therefore exploit this result to design a new frequency domain test statistic, which is shown to be consistent in the high-dimensional regime. This result is connected to the widely studied \emph{spiked models} in Random Matrix Theory, defined as low rank perturbations of large random matrices. These models were extensively studied in the context of sample covariance matrices of independent identically distributed high-dimensional vectors, see e.g. \cite{Baik2006}. We however notice that papers addressing the behaviour of the corresponding sample correlation matrices are quite scarce, see  \cite{Morales2019} when the low rank perturbation affects only the first components of the observations.

\subsection{Related works}
Although the asymptotic framework differs from the high-dimensional regime considered here, we also mention the series of studies \cite{Forni2000, Forni2004} in the econometrics field, which consider a similar model under $\Hcal_1$. In these works, it is assumed that $M,N \to \infty$ so the ratio $\frac{M}{N}$ remains bounded, while the $K$ non-zero eigenvalues of the spectral density $\S_{\u}(\nu)$ of $(\u_n)_{n \in \Zbb}$ are assumed to converge towards $+\infty$ at rate $M$. This last assumption, which ensures that the Signal-to-Noise Ratio (SNR)
$\frac{\Ebb \|\u_n\|^2}{\Ebb\|\v_n\|^2}$
remains bounded away from $0$ as $M \to\infty$, significantly facilitates the design of consistent detection methods.
Nevertheless, while relevant in the domain of econometrics, this assumption may be unrealistic in several applications of array processing, where the challenge is to manage situations in which the SNR
converges towards $0$ at rate $\frac{1}{M}$.
This situation is the one considered in this paper and, in that case, the results of \cite{Forni2000, Forni2004} can not be used. We discuss this point further in Section \ref{section:model} below.

The rest of the paper is organized as follows. In Section \ref{section:model}, we formally introduce the model of signals used in the remainder, as well as the required technical assumptions. In section \ref{section:test_statistics_introduction}, we introduce informally the proposed test statistic, and illustrate its behaviour in order to provide some intuition before a more rigorous presentation. In section \ref{section:scm}, we study some approximation results for the spectral coherence $\hat{\C}_{\y}(\nu)$ which are useful to study the linear spectral statistics considered here. This study is then used in Section \ref{section:test} to introduce a new test statistic that is consistent in the high-dimensional regime. Finally Section \ref{section:simulations} provides some simulations illustrating its performance and comparisons against other relevant approaches.

\textit{Notations.} For a complex matrix $\A$, we denote by $\A^*$ its conjugate transpose, and by $\|\A\|_2$ and $\|\A\|_F$ its spectral and Frobenius norms respectively. If $\A$ is a $n \times n$ complex matrix, we denote by $\Tr(\A)$ its trace, and by $\lambda_1(\A),\ldots,\lambda_n(\A)$ its eigenvalues; if moreover $\A$ is Hermitian, they are sorted in decreasing order $\lambda_1(\A) \geq \ldots \geq \lambda_n(\A)$. The $n \times n$ identity matrix is denoted $\I_n$.
The expectation of a complex random variable $Z$ is denoted by $\Ebb[Z]$.
The complex circular Gaussian distribution with variance $\sigma^2$ is denoted as $\Ncal_{\Cbb}(0,\sigma^2)$ and a random vector $\x$ of $\Cbb^n$ follows the $\Ncal_{\Cbb^n}(\mathbf{0},\mathbf{\R})$ distribution if $\b^*\x \sim \Ncal_{\Cbb}(0,\b^*\R\b)$ for all deterministic (column) vector $\b$ and a fixed $n \times n$ positive definite matrix $\R$.
Finally, $\Ccal^1(I)$ (resp. $\Ccal^1_c(I)$) represents the set of continuously differentiable functions (resp. continuously differentiable functions with compact support) on an open set $I$. 

\section{Model and assumptions}
\label{section:model}

Let us consider a $M$--dimensional observed time series $(\y_n)_{n \in \Zbb}$ defined as
\begin{align}
  \y_n = \u_n + \v_n
  \label{eq:model}
\end{align}
where $(\u_n)_{n \in \Zbb}$ represents a useful signal and where $(\v_n)_{n \in \Zbb}$ represents an additive noise. The useful signal is modeled as the output of an unknown stable MIMO filter $(\H_k)_{k \in \Zbb}$
driven by a non-observable $K$--dimensional complex Gaussian white noise $(\epsilonbs_n)_{n \in \Zbb}$ with $\Ebb[\epsilonbs_n\epsilonbs_n^*] = \I_K$, i.e.
\begin{align}
  \u_n = \sum_{k \in \Zbb} \H_k \epsilonbs_{n-k}
  \notag
\end{align}
with probability one. We notice that 
$K$ represents the number of sources in the context of array processing. $(\v_n)_{n \in \Zbb}$ is modeled as a $M$--dimensional stationary complex Gaussian time series such that its component time series $(v_{1,n})_{n \in \Zbb},\ldots,(v_{M,n})_{n \in \Zbb}$ are mutually independent. 

For each $m = 1 ,\ldots,M$, we denote by $(r_m(k))_{k \in \Zbb}$ the covariance function of $(v_{m,n})_{n \in \Zbb}$, i.e.
$r_m(k) = \Ebb[v_{m,n} \overline{v_{m,n-k}}]$, which verifies the following memory assumption.
\begin{assumption}
  \label{assumption:rm}
  The covariance coefficients decay sufficiently fast in the lag domain, in the sense that
  \begin{align}
  \label{eq:assumption-rm}
    \sup_{m \geq 1} \sum_{k \in \Zbb} (1+|k|)^2 |r_m(k)| < \infty.
  \end{align}
\end{assumption}
In particular, Assumption \ref{assumption:rm} implies that the spectral density $s_m$ of $(v_{m,n})_{n \in \Zbb}$, given by
\begin{align}
  s_m(\nu) = \sum_{k \in \Zbb} r_m(k) \erm^{-\irm 2 \pi \nu k}
  \notag
\end{align}
verifies
\begin{align}
  \sup_{m \geq 1} \sup_{\nu \in [0,1]} s_m(\nu) < \infty.
  \notag
\end{align}
Assumption \ref{assumption:rm} is in particular verified as soon as the condition 
\begin{equation}
    \label{eq:rate-convergence-autocovariance}
|r_m(k)| \leq \frac{C}{(1+|k|)^{3+\delta}} 
\end{equation}
holds for each $k \in \mathbb{Z}$ and each $m \geq 1$, where $C$ and $\delta$ are positive constants. As the autocovariance 
function of ARMA signals decreases exponentially towards $0$, Assumption \ref{assumption:rm} thus holds if the time series 
$(v_m)_{m \geq 1}$ are ARMA signals, provided some extra purely technical conditions that allow to manage the supremum over $m$ in 
(\ref{eq:assumption-rm}) are met. As the spectral coherence matrix of $(\v_n)_{n \in \Zbb}$, involves a renormalization by the inverse of the spectral densities $s_m$, we also need that $s_m$ does not vanish for each $m$.
\begin{assumption}
  \label{assumption:sm}
  The spectral densities are uniformly bounded away from zero, that is
  \begin{align}
    \inf_{m \geq 1}  \min_{\nu \in [0,1]} s_m(\nu) > 0.
    \notag
  \end{align}
\end{assumption}
Assumptions \ref{assumption:rm} and \ref{assumption:sm} also imply that the total noise power satisfies
\begin{align}
\label{eq:noise-power-bounds}
  0 < \inf_{M \geq 1} \frac{1}{M} \Ebb \|\v_n\|_2^2 \leq \sup_{M \geq 1} \frac{1}{M} \Ebb \|\v_n\|_2^2 < \infty.
\end{align}
The next assumption is related to the signal part $(\u_n)_{n \in \Zbb}$. For each $\nu \in [0,1]$, we denote by $\H(\nu)$ the Fourier transform of $(\H_k)_{k \in \Zbb}$, i.e.
\begin{align}
  \H(\nu) = \sum_{k \in \Zbb} \H_k \erm^{-\irm 2 \pi \nu k}
  \notag
\end{align}
and by $\h^1(\nu),\ldots,\h^M(\nu)$ the rows of $\H(\nu)$.
\begin{assumption}
  \label{assumption:H}
  The MIMO filter coefficient matrices are such that
  \begin{align}
    \sup_{M \geq 1} \sum_{k \in \Zbb} (1+|k|)\left\|\H_k\right\|_2 < \infty
    \label{assumption:memory_H}
  \end{align}
  and
  \begin{align}
    \lim_{M \to \infty} \max_{m=1,\ldots,M} \max_{\nu \in [0,1]} \left\|\h^m(\nu)\right\|_2 = 0.
    \label{assumption:power_H}
  \end{align}
\end{assumption}
When $K$ is fixed while $M \to \infty$, condition \eqref{assumption:memory_H} in Assumption \ref{assumption:H} implies that the total useful signal power remains bounded, i.e.
\begin{align}
  \Ebb \left\|\u_n\right\|_2^2 = \sum_{k \in \Zbb} \left\|\H_k\right\|_F^2 = \Ocal(1)
  \label{eq:condition-snr}
\end{align}
so that, using (\ref{eq:noise-power-bounds}), the SNR
vanishes at rate $\frac{1}{M}$, i.e.
\begin{align}
  \frac{\Ebb\|\u_n\|_2^2}{\Ebb\|\v_n\|_2^2} = \Ocal\left(\frac{1}{M}\right).
  \label{eq:SNR}
\end{align}
Likewise, condition \eqref{assumption:power_H} in Assumption \ref{assumption:H} implies that the SNR per time series vanishes, i.e.
\begin{align}
  \frac{\Ebb|u_{m,n}|^2}{\Ebb|v_{m,n}|^2} = \frac{\int_0^1 \|\h^m(\nu)\|_2^2 \drm \nu}{\int_0^1 s_m(\nu) \drm \nu} = o(1)
  \label{eq:SNR_per_ts}
\end{align}
as $M \to \infty$. We finally notice that (\ref{assumption:memory_H}) is stronger than (\ref{eq:condition-snr}). While $\Ebb \left\|\u_n\right\|_2^2 = \mathcal{O}(1)$ is a rather fundamental assumption that allows to precise the behaviour of the signal to noise ratio,  the extra condition $\sup_{m\ge1}\sum_{k} |k| \left\|\H_k\right\|_2 < \infty$ is essentially motivated by technical reasons (it is needed to establish Theorem \ref{theorem:pure_signal}). However, it is clearly not restrictive in practice.
\begin{remark}
  Conditions \eqref{assumption:memory_H} and \eqref{assumption:power_H} in Assumption \ref{assumption:H} are especially relevant in the context of array processing,
  where $M$ represents the number of sensors, which may be large \cite{Vallet2012, Vallet2015}. In this context, \eqref{eq:SNR} represents the SNR before matched filtering, while \eqref{eq:SNR_per_ts}
  represents the SNR per sensor. The use of spatial filtering techniques, which combine the observations $y_{1,n},\ldots,y_{M,n}$ across the $M$ sensors, allows to increase the SNR
  by a factor $M$ when the second order statistics of $(\y_n)_{n \in \Zbb}$ are known, which leads to a SNR after matched filtering of the order of magnitude $\Ocal(1)$. 
  Thus, despite the apparent low SNR, reliable information on the useful signal $(\u_n)_{n \in \Zbb}$ can potentially still be extracted from the observed signal $(\y_n)_{n \in \Zbb}$.
\end{remark}
Let $\S_{\y}$ denote the spectral density of $(\y_n)_{n \in \Zbb}$, given by
\begin{align}
  \S_{\y}(\nu) = \H(\nu)\H(\nu)^* + \S_{\v}(\nu)
  \notag
\end{align}
where $\S_{\v}(\nu) = \dg\left(s_1(\nu),\ldots,s_M(\nu)\right)$. To estimate $\S_{\y}$, we consider in this paper a frequency-smoothed periodogram $\hat{\S}_{\y}$, which we defined in \eqref{equation:definition_S_y}. In the classical low-dimensional regime where $B,N \to \infty$ while $M,K$ remain fixed, it is well-known
\cite{Brillinger1981} that
\begin{align}
  \Ebb[\hat{\S}_{\y}(\nu)] = \S_{\y}(\nu) + \Ocal\left(\frac{B^2}{N^2}\right)
  \notag
\end{align}
and
\begin{align}
  \Ebb\left\|\hat{\S}_{\y}(\nu) - \Ebb[\hat{\S}_{\y}(\nu)]\right\|_2^2 = \Ocal\left(\frac{1}{B}\right).
  \notag
\end{align}
Thus, in this regime, $\hat{\S}_{\y}(\nu)$ is a consistent estimator of $\S_{\y}(\nu)$ as long as $B \to \infty$ and $\frac{B}{N} \to 0$.
Likewise, the sample Spectral Coherence Matrix (SCM, not to be confused with the sample covariance matrix, which will not be used in this paper) defined in \eqref{equation:definition_hat_C_y} is a consistent estimator of the true SCM $\C_{\y}(\nu)$ defined in \eqref{equation:definition_C_y}.
When $M \rightarrow +\infty$ and $\frac{M}{N} \rightarrow 0$, it can be shown that, under some additional mild extra assumptions, the consistency of $\hat{\S}_{\y}(\nu)$ and  $\hat{\C}_{\y}(\nu)$ in the spectral norm sense still holds provided that $B$ is chosen in such a way that $\frac{B}{N} \rightarrow 0$ and $\frac{M}{B} \rightarrow 0$. In practice, for finite values of $M$ and $N$, the above asymptotic regime will allow to predict the performance of various inference schemes in situations where it is possible to choose $B$ in such a way that $\frac{M}{B}$ and $\frac{B}{N}$ are both small enough.
Nevertheless, when the dimension $M$ is large and the sample size $N$ is not unlimited, or equivalently if $\frac{M}{N}$ is not small enough, such a choice of $B$ may be
impossible. In such a context, it seems more relevant to consider asymptotic regimes for which $\frac{M}{N} \rightarrow 0$ and $\frac{M}{B}$ converging towards a positive constant. In the following, we will consider the following asymptotic regime. 
\begin{assumption}
  \label{assumption:regime}
  $N = N(M)$ and $B = B(M)$ are both functions of $M$ such that, for some $\alpha \in (0,1)$,
  \begin{align}
    M = \Ocal\left(N^{\alpha}\right) \text{ and } \frac{M}{B} \xrightarrow[M\to\infty]{} c \in (0,1)
    \notag
  \end{align}
  while $K$ is fixed with respect to $M$.
\end{assumption}
As $\frac{M}{B}$ does not converge towards $0$, the consistency of $\hat{\S}_{\y}(\nu)$ and  $\hat{\C}_{\y}(\nu)$ is lost. This can be explained in a simple way when $\u_n = 0$ for each $n$ and the signals $((v_{m,n})_{n \in \mathbb{Z}})_{m \geq 1}$ are mutually independent i.i.d. $\mathcal{N}_c(0,\sigma^{2})$ distributed sequences. In this context, for each $\nu$, the renormalized Fourier transform vectors 
$(\xibs_{\y}(\nu +b/N))_{b=-B/2, \ldots, B/2}$ are mutually independent $\mathcal{N}_{\mathbb{C}}(0, \sigma^{2} \I)$ random vectors. 
The spectral density estimate $\hat{\S}_{\y}(\nu)$ defined by (\ref{equation:definition_S_y}) thus coincides with the sample covariance matrix of these $(B+1)$ $M$--dimensional vectors. If $B$ and $M$ are of the same order to magnitude, it cannot be expected that $\| \hat{\S}_{\y}(\nu) - \mathbb{E}(\hat{\S}_{\y}(\nu)) \|$ converges towards $0$ because the true covariance matrix $\mathbb{E}(\hat{\S}_{\y}(\nu))$ to be estimated depends on $\mathcal{O}(M^{2})$ 
parameters, and that the number $MB$ of available scalar observations used to estimate $\mathbb{E}(\hat{\S}_{\y}(\nu))$ is also $\mathcal{O}(M^{2})$. 
Despite the loss of the convergence of the estimators  $\hat{\S}_{\y}(\nu)$ and  $\hat{\C}_{\y}(\nu)$,  we will see that one can still rely on the high-dimensional structure of these matrices to design relevant test statistics.

\section{Informal presentation of the proposed test statistic}
\label{section:test_statistics_introduction}
Mathematical details will reveal later that for each $\nu$, $\hat{\C}(\nu)$ behaves as a spike model covariance matrix, whose eigenvalues are precisely described by \cite{Baik2006}. More precisely, we will see that, in some sense, the eigenvalues of $\hat{\C}(\nu)$ that are due to the noise belong to the interval $[\lambda_{-}, \lambda_{+}]$ where $\lambda_{-} = (1-\sqrt{c})^2$ 
and $\lambda_{+} = (1+\sqrt{c})^2$, and that in the presence of signal, some eigenvalues of $\hat{\C}(\nu)$ may be strictly greater than $\lambda_+$ if an SNR criteria is respected.
For the remainder, we define
\begin{align}
  \Vcal_N = \left\{\frac{k}{N} : k=0,\ldots,N-1\right\}
  \label{eq:V_set}
\end{align}
the set of Fourier frequencies. A natural way to test for $\Hcal_0$ against $\Hcal_1$ is to compute the largest eigenvalue of $\hat{\C}(\nu)$ over the frequencies of $\Vcal_N$, and compare it with $\lambda_+$. This leads to the following test statistic:
\begin{align}
  T_{\epsilon} = \mathbb{1}_{[\lambda^+ + \epsilon,\infty)}\left(\max_{\nu \in \Vcal_N} \lambda_1\left(\hat{\C}_{\y}(\nu)\right)\right).
  \label{eq:LE}
\end{align}
We will prove later that, under proper assumption on the SNR, this test statistic is consistent in the present high-dimensional regime. Before describing the mathematical details leading to consider $T_\epsilon$, we now provide some numerical illustrations of its behaviour. The general settings are given as follows. The noise is generated
as a Gaussian AR(1) process having spectral density
\begin{align}
\label{equation:s_m_simulations}
  s_m(\nu) = \frac{1}{\left|1-\theta\erm^{-\irm 2 \pi \nu}\right|^2},
\end{align}
for all $m=1,\ldots,M$, with $\theta=0.5$,
whereas for the useful signal, we also consider an AR(1) process by choosing $K=1$ and
\begin{equation}
\label{equation:H_k_simulations}
    \H_k =  \sqrt{\frac{C}{M}} \beta^k (1,\ldots,1)^T
\end{equation}
with $\beta = \frac{10}{11}$ and $C$ being a positive constant used to adjust the SNR. 

\begin{figure}[!h]
\includegraphics[width=\columnwidth]{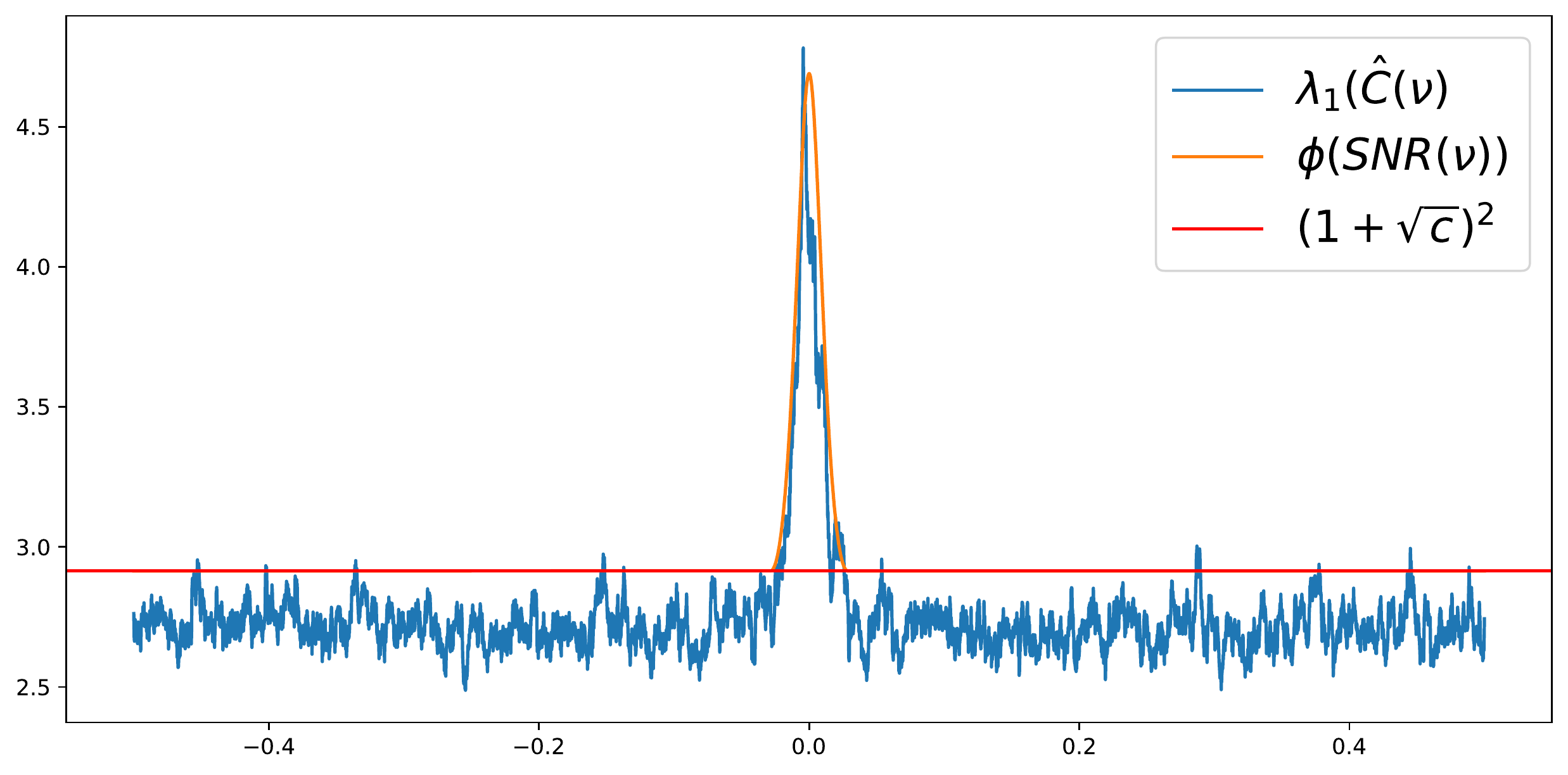}
\caption{Largest eigenvalue of $\hat{\C}_{\y}(\nu)$ for $\nu\in\Vcal_N$ vs the threshold $\lambda_+=(1+\sqrt{\frac{M}{B+1}})^2$. $M=60$, $c=0.5$, $N=6000$, $\theta=0.5$, $C=0.05$} 
\label{figure:eigenvalues_over_frequencies} 
\end{figure}

\begin{figure}[!h]
\includegraphics[width=\columnwidth]{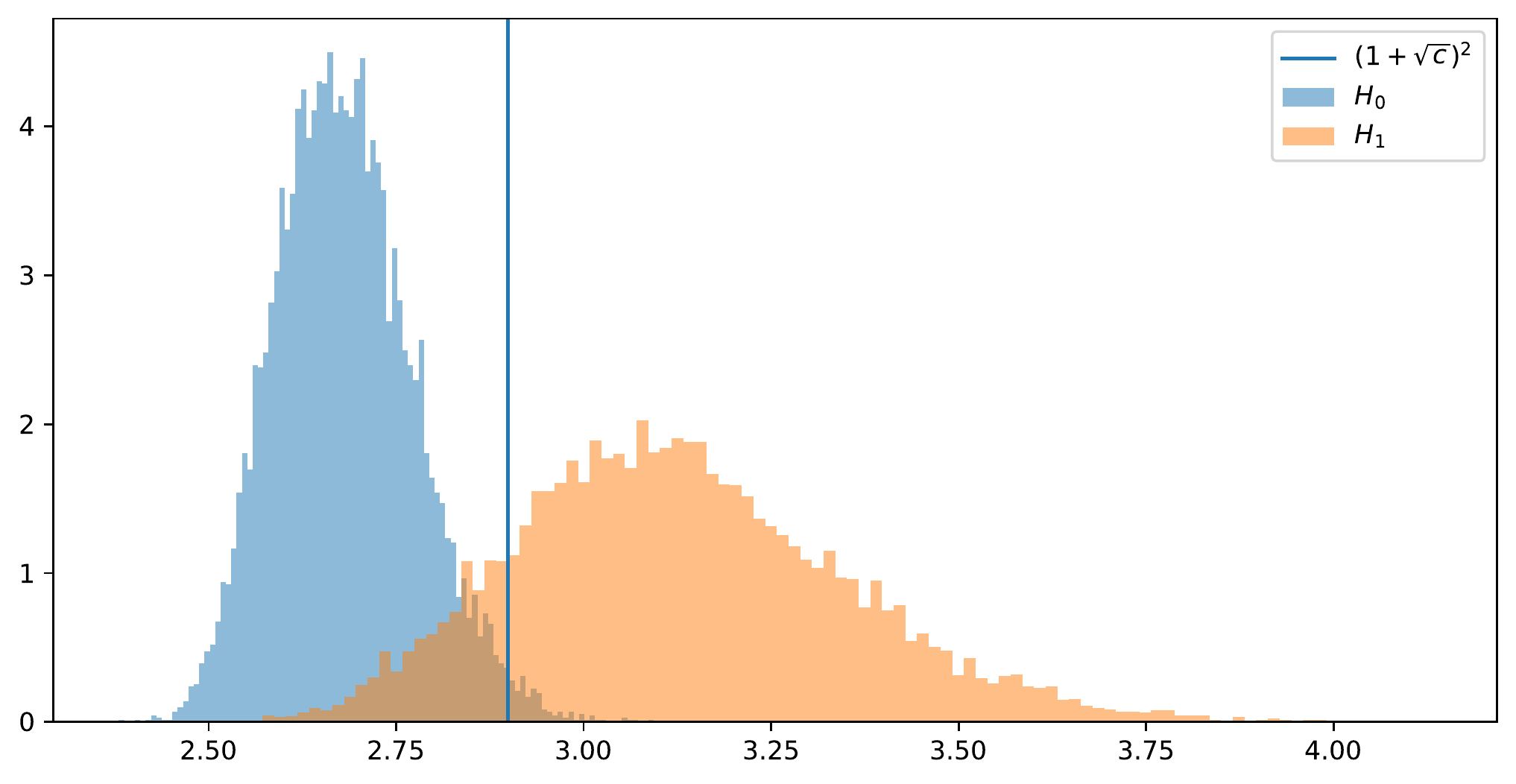}
\caption{Histogram of $T_\epsilon$ under $\Hcal_0$ and $\Hcal_1$, over 10000 repetitions. $M=40$, $c=0.5$, $N=1000$, $\theta=0.5$, $C=0.05$} 
\label{figure:histogram_eigenvalues} 
\end{figure}

In order to understand how the test statistics $T_\epsilon$ discriminates between $\Hcal_0$ and $\Hcal_1$, we show in Figure \ref{figure:eigenvalues_over_frequencies} the largest eigenvalue of $\hat{\C}_\y(\nu)$ for $\nu\in\Vcal_N$ in the presence of signal, and compare it to the threshold $\lambda_+$. We see that for some frequencies $\nu$ around $0$, the largest eigenvalue of $\hat{\C}_\y(\nu)$ deviates significantly from $\lambda_+$. As we will see later, it is possible to 
evaluate the asymptotic behaviour of the largest eigenvalue of $\hat{\C}_\y(\nu)$, and to establish that it converges towards 
$\phi(SNR(\nu))$ where $\phi$ is a certain function, and where $SNR(\nu)$ can be interpreted as a signal-to-noise ratio at frequency $\nu$. $\phi(SNR(\nu))$ is also represented in Figure \ref{figure:eigenvalues_over_frequencies}, and it is seen that it is close to the largest eigenvalue of $\hat{\C}_\y(\nu)$. In Figure \ref{figure:histogram_eigenvalues}, we compare the empirical distribution of $T_\epsilon$ under $\Hcal_0$ and $\Hcal_1$ over 10000 repetitions. We see that the distribution of our test statistic $T_\epsilon$ is able to discriminate the scenarios where the data $\y_n$ are generated under $\Hcal_0$ or $\Hcal_1$, and that $T_\epsilon$ is typically over the threshold $\lambda_+$ under $\Hcal_1$.

\section{Approximation results for $\hat{\C}_{\y}(\nu)$ in the high-dimensional regime}
\label{section:scm}

In this section we present the mathematical details which lead to the test statistic \eqref{eq:LE}. More specifically, 
we provide useful approximation results for $\hat{\C}_{\y}(\nu)$, which basically show that $\hat{\C}_{\y}(\nu)$ behaves as a certain Wishart matrix in the high-dimensional regime.
These approximation results are the keystone for the study of the behaviour of the eigenvalues of $\hat{\C}_{\y}(\nu)$ and the detection test proposed in Section \ref{section:test}.

We first study separately the signal-free case (i.e. $\y_n = \v_n$) as well as the noise-free case (i.e. $\y_n = \u_n$). 

\subsection{Signal-free case}
Let 
$$ \xibs_{\v}(\nu) = \frac{1}{\sqrt{N}} \sum_{n=0}^{N-1} \v_n \erm^{-\irm 2 \pi \nu n}$$
denote the discrete (time-limited) Fourier transform of $(\v_n)_{n=0,\ldots,N-1}$, and define the $M \times (B+1)$ matrix $\Sigmabs_{\v}(\nu)$ as
\begin{align}
  \Sigmabs_{\v}(\nu) = \frac{1}{\sqrt{B+1}}\left[\xibs_{\v}\left(\nu-\frac{B}{2N}\right),\ldots,\xibs_{\v}\left(\nu+\frac{B}{2N}\right)\right].
  \notag
\end{align}
The following result, derived in \cite{loubaton2020large}, reveals an interesting behaviour of the frequency-smoothed periodogram of the noise.
\begin{theorem} 
  \label{theorem:pure_noise}
  Under Assumptions \ref{assumption:rm}, \ref{assumption:sm}  and \ref{assumption:regime}, for all $\nu \in \Vcal_N$,
  there exists an $M\times(B+1)$ matrix $\Z(\nu)$ with i.i.d. $\Ncal_{\Cbb}(0,1)$ entries such that
  \begin{align}
    \max_{\nu\in  \Vcal_N} \left\|\Sigmabs_{\v}(\nu) - \frac{1}{\sqrt{B+1}} \S_{\v}(\nu)^{1/2} \Z(\nu)\right\|_2
    \xrightarrow[M\to\infty]{a.s.} 0.
    \notag
  \end{align}
\end{theorem}
Informally speaking, Theorem \ref{theorem:pure_noise} shows that the random vectors
$\frac{1}{\sqrt{B+1}}\xibs_{\v}\left(\nu-\frac{B}{N}\right)$,\ldots,$\frac{1}{\sqrt{B+1}}\xibs_{\v}\left(\nu+\frac{B}{N}\right)$
asymptotically behave as a family of i.i.d. $\Ncal_{\Cbb^M}(\mathbf{0},\S_{\v}(\nu))$ vectors, for all $\nu \in \Vcal_N$.
Moreover, if
\begin{align}
  \hat{\S}_{\v}(\nu) := \frac{1}{B+1} \sum_{b=-B/2}^{B/2} \xibs_{\v}\left(\nu + \frac{b}{N}\right) \xibs_{\v}\left(\nu + \frac{b}{N}\right)^*
  \notag
\end{align}
denotes the frequency-smoothed periodogram of the noise observations $(\v_n)_{n \in \Zbb}$, we deduce that $\hat{\S}_{\v}(\nu)$ asymptotically behaves as a complex Gaussian Wishart matrix with covariance matrix $\S_{\v}(\nu)$, thanks to the following corollary.
\begin{corollary}
  \label{corollary:pure_noise}
  Under the assumptions of Theorem \ref{theorem:pure_noise}, it holds that
  \begin{align}
    \max_{\nu \in \Vcal_N} \left\|\hat{\S}_{\v}(\nu) - \S_{\v}(\nu)^{1/2} \frac{\Z(\nu)\Z(\nu)^*}{B+1}\S_{\v}(\nu)^{1/2}\right\|_2
    \xrightarrow[M\to\infty]{a.s.} 0.
    \notag
  \end{align}
\end{corollary}
\begin{IEEEproof}
  The proof is deferred to Appendix \ref{appendix:proof_corollary_pure_noise}.
\end{IEEEproof}
It is worth noticing that Corollary \ref{corollary:pure_noise} implies in particular
\begin{align}
 \max_{\nu \in \Vcal_N} \left\|\dg\left(\hat{\S}_{\v}(\nu)\right) - \S_{\v}(\nu)\right\|
  \xrightarrow[M\to\infty]{a.s.} 0
  \notag
\end{align}
and consequently $\dg(\hat{\S}(\nu))$ is a consistent estimator of the noise spectral density $\S_{\v}(\nu)$ in the operator norm sense,
at each Fourier frequency $\nu \in \Vcal_N$. This convergence may be directly obtained using Lemma \ref{lemma:hanson-wright} in
Appendix \ref{appendix:useful_results} and we omit the details since this result is well-known.


\subsection{Noise-free case}

Let $$\xibs_{\u}(\nu) = \frac{1}{\sqrt{N}} \sum_{n=0}^{N-1} \u_n \erm^{-\irm 2 \pi \nu n}$$ and let $\Sigmabs_{\u}(\nu)$ be the $K \times (B+1)$ matrix defined as
\begin{align}
  \Sigmabs_{\u}(\nu) = \frac{1}{\sqrt{B+1}} \left[\xibs_{\u}\left(\nu - \frac{B}{2N}\right),\ldots,\xibs_{\u}\left(\nu + \frac{B}{2N}\right)\right].
  \notag
\end{align}
In the same way, we also denote by  $\xibs_{\epsilonbs}$ the normalized discrete (time-limited) Fourier transform of $(\epsilonbs_n)_{n=0,\ldots,N-1}$, and consider the $K \times (B+1)$ matrix
$\Sigmabs_{\epsilonbs}(\nu)$ defined as $\Sigmabs_{\u}(\nu)$. We then have the following important approximation result.
\begin{theorem}
  \label{theorem:pure_signal}
  Under Assumptions \ref{assumption:H} and \ref{assumption:regime}, it holds that
  \begin{align}
    \max_{\nu \in \Vcal_N} \left\|\Sigmabs_{\u}(\nu) - \H(\nu)\Sigmabs_{\epsilonbs}(\nu)\right\|_2
    \xrightarrow[M\to\infty]{a.s.} 0.
    \notag
  \end{align}
\end{theorem}
\begin{IEEEproof}
  The proof is deferred to Appendix \ref{appendix:proof_theorem_pure_signal}.  
\end{IEEEproof}
As in Theorem \ref{theorem:pure_noise}, Theorem \ref{theorem:pure_signal} shows that the random vectors
$\xibs_{\u}\left(\nu-\frac{B}{N}\right),\ldots,\xibs_{\u}\left(\nu+\frac{B}{N}\right)$ asymptotically behave as the i.i.d. vectors
$\H(\nu)\xibs_{\epsilonbs}\left(\nu-\frac{B}{N}\right),\ldots,\H(\nu)\xibs_{\epsilonbs}\left(\nu+\frac{B}{N}\right)$, for all $\nu \in \Vcal_N$.
\begin{remark}
  The type of approximation given in Theorem \ref{theorem:pure_signal} is well-known in the low-dimensional regime in which $M,K,B$
  are fixed while $N\to\infty$. Indeed, in that case, we have \cite[Th. 4.5.2]{Brillinger1981}
  \begin{align}
    \max_{\nu \in [0,1]} \left\|\Sigmabs_{\u}(\nu) - \H(\nu) \Sigmabs_{\epsilonbs}(\nu)\right\|_2
    = \Ocal_P\left(\sqrt{\frac{\log(N)}{N}}\right).
    \notag
  \end{align}
  In the high-dimensional regime where $M$ and $B$ also converge to infinity as described in Assumption \ref{assumption:regime},
  the result of Theorem \ref{theorem:pure_signal} cannot be obtained from \cite[Th. 4.5.2]{Brillinger1981} and thus requires a new study.
\end{remark}
We also deduce the following approximation result on the frequency-smoothed periodogram of the signal observations $(\u_n)_{n=0,\ldots,N-1}$ given by
\begin{align}
    \hat{\S}_{\u}(\nu) :=
    \frac{1}{B+1} \sum_{b=-B/2}^{B/2} \xibs_{\u}\left(\nu + \frac{b}{N}\right) \xibs_{\u}\left(\nu + \frac{b}{N}\right)^*.
  \notag
\end{align}
\begin{corollary}
  \label{corollary:pure_signal}
  Under the assumptions of Theorem \ref{theorem:pure_signal}, it holds that
  \begin{align}
    \max_{\nu \in \Vcal_N} \left\|\hat{\S}_{\u}(\nu) - \H(\nu)\H(\nu)^*\right\|_2
    \xrightarrow[M\to\infty]{a.s.} 0.
    \notag
  \end{align}
\end{corollary}
\begin{IEEEproof}
  The proof is deferred to Appendix \ref{appendix:proof_corollary_pure_signal}.
\end{IEEEproof}
As a result of Corollary \ref{corollary:pure_signal}, we deduce that the frequency-smoothed periodogram $\hat{\S}_{\u}(\nu)$
is a consistent estimator of the spectral density $\S_{\u}(\nu) = \H(\nu)\H(\nu)^*$ of $(\u_n)_{n \in \Zbb}$ in the high-dimensional regime, for each $\nu \in \Vcal_N$.

Having characterized the pure noise and pure signal cases, we are now in position to study the high-dimensional behaviour of the spectral coherence matrix $\hat{\C}_{\y}(\nu)$.

\subsection{The signal-plus-noise case}

First, using Corollaries \ref{corollary:pure_noise} and \ref{corollary:pure_signal},
we deduce the high-dimensional behaviour of the frequency smoothed periodogram $\hat{\S}_{\y}(\nu)$. The following results show that, as it could be expected, the frequency smoothed periodogram essentially behaves as a colored Wishart matrix in the large asymptotic regime.
\begin{proposition}
   \label{proposition:S_hat_y}
  For all $\nu \in \Vcal_N$, there exists an $M \times (B+1)$ matrix $\X(\nu)$ with i.i.d. $\Ncal_{\Cbb}(0,1)$ entries such that
  \begin{align}
    \max_{\nu \in \Vcal_N}
    \left\|\hat{\S}_{\y}(\nu) - \S_{\y}(\nu)^{\frac{1}{2}}\frac{\X(\nu)\X(\nu)^*}{B+1}\S_{\y}(\nu)^{\frac{1}{2}}\right\|_2
    \xrightarrow[M\to\infty]{a.s.} 0.
    \label{eq:conv_hat_S_y}
  \end{align}
\end{proposition}
\begin{IEEEproof}
  The proof is deferred to Appendix \ref{appendix:proof_proposition_S_hat_y}.
\end{IEEEproof}
We finally consider the study of the spectral coherence
$\hat{\C}_{\y}(\nu) = \dg(\hat{\S}_{\y}(\nu))^{-\frac{1}{2}} \hat{\S}_{\y}(\nu) \dg(\hat{\S}_{\y}(\nu)^{-\frac{1}{2}}$.
From condition \eqref{assumption:power_H} in Assumption \ref{assumption:H} on the SNR, it turns out that (cf. proof of Theorem \ref{theorem:C_hat_y} below where the result is shown) that
\begin{align}
    \label{equation:approximation_S_v_diag_S_y}
   \max_{\nu \in \Vcal_N} \left\|\dg\left(\hat{\S}_{\y}(\nu)\right) - \S_{\v}(\nu)\right\|_2
  \xrightarrow[M\to\infty]{a.s.} 0.
\end{align}
This approximation result regarding the normalization term  $\dg(\hat{\S}_{\y}(\nu))$ in the SCM  naturally leads to the following theorem, which is the key result of this paper.
\begin{theorem}
  \label{theorem:C_hat_y}
  Under Assumptions \ref{assumption:rm}, \ref{assumption:sm}, \ref{assumption:H} and \ref{assumption:regime}, 
  \begin{align}
    \max_{\nu \in \Vcal_N}
    \left\|
    \hat{\C}_{\y}(\nu)
    -
    \Xibs(\nu)^{\frac{1}{2}} \frac{\X(\nu)\X(\nu)^*}{B+1}\Xibs(\nu)^{\frac{1}{2}}\right\|_2
    \xrightarrow[M\to\infty]{a.s.} 0 
    \notag
  \end{align}
  where
  \begin{align}
    \Xibs(\nu) = \S_{\v}(\nu)^{-\frac{1}{2}} \H(\nu)\H(\nu)^*\S_{\v}(\nu)^{-\frac{1}{2}} + \I_M.
    \notag
  \end{align}
  and $\X(\nu)$ is the matrix defined in Proposition \ref{proposition:S_hat_y}.
\end{theorem}
\begin{IEEEproof}
 The proof is deferred to Appendix \ref{appendix:proof_theorem_SCM}. 
\end{IEEEproof}
Let us make a few important comments regarding the result of Theorem \ref{theorem:C_hat_y}.

First, used in conjunction with Weyl's inequalities \cite[Th. 4.3.1]{Horn2005}, Theorem \ref{theorem:C_hat_y} implies in particular that each eigenvalue of the SCM $\hat{\C}_{\y}(\nu)$ behaves as its counterpart
of the Wishart matrix $$\W(\nu) = \Xibs(\nu)^{\frac{1}{2}} \frac{\X(\nu)\X(\nu)^*}{B+1}\Xibs(\nu)^{\frac{1}{2}}$$ for $\nu \in \Vcal_N$, that is
\begin{align}
  \max_{m=1,\ldots,M} \max_{\nu \in \Vcal_N}
  \Bigl|
  \lambda_m\left(\hat{\C}_{\y}(\nu)\right)
    -
  \lambda_m\left(\W(\nu)\right)
  \Bigr|
  \xrightarrow[M\to\infty]{a.s.} 0.
  \label{eq:conv_eig}
\end{align}
Second, Theorem \ref{theorem:C_hat_y} has an important consequence regarding the behaviour of linear spectral statistics of $\hat{\C}_{\y}(\nu)$, that is statistics of the type
\begin{align}
  L_{\varphi}(\nu) = \frac{1}{M} \sum_{m=1}^M \varphi\left(\lambda_m\left(\hat{\C}_{\y}(\nu)\right)\right)
  \label{eq:lss}
\end{align}
where $\varphi$ belongs to a certain class of functions.
\begin{corollary}
  \label{corollary:lss}
  Let $\varphi \in \Ccal^1\left((0,+\infty)\right)$.
  Under Assumptions \ref{assumption:rm}, \ref{assumption:sm}, \ref{assumption:H} and \ref{assumption:regime}, we have
  \begin{align}
    \max_{\nu \in \Vcal_N}
    \left|
    L_{\varphi}(\nu)
    - \int_{\Rbb} \varphi(\lambda) f(\lambda)\drm\lambda
    \right|
    \xrightarrow[M\to\infty]{a.s.} 0
    \notag
  \end{align}
   where $f$ is the density of the Marcenko-Pastur distribution given by
  \begin{align}
    f(\lambda) =
    \frac{\sqrt{(\lambda-\lambda^-)(\lambda^+ -\lambda)}}{2 \pi c \lambda} \mathbb{1}_{[\lambda^-,\lambda^+]}(\lambda)
\notag
  \end{align}
  with $\lambda^{\pm} = \left(1\pm\sqrt{c}\right)^2$.
\end{corollary}
\begin{IEEEproof}
  The proof is deferred to Appendix \ref{appendix:proof_corollary_lss}.
\end{IEEEproof}
Therefore, Corollary \ref{corollary:lss} shows that linear spectral statistics of the SCM converge to the same limit regardless of whether the observations contain only pure noise or signal-plus-noise contributions. This shows that any test statistic solely relying on linear spectral statistics of the SCM is unable to distinguish between absence or presence of useful signal, and cannot be consistent in the high-dimensional regime.
Nevertheless, in the next section we will see that we can exploit Theorem \ref{theorem:C_hat_y} to build a new test statistic based on the largest eigenvalue of $\hat{\C}_{\y}(\nu)$, which is proved to be consistent in the high-dimensional regime.

\begin{remark}
  Corollary \ref{corollary:pure_noise}, Corollary \ref{corollary:pure_signal} and Theorems \ref{theorem:C_hat_y} may be interpreted in the context of array processing.
  Indeed, in the time model \eqref{eq:model}, usually referred to as ``wideband'', the signal contribution $(\u_n)_{n \in \Zbb}$ modeled as a linear process, is in general not confined to a low-dimensional subspace (i.e. with dimension less than $M$).  However, in the frequency domain, Corollary \ref{corollary:pure_noise} and Corollary \ref{corollary:pure_signal} show that we can retrieve, in the high-dimensional regime, a ``narrowband'' model, since the useful signal is confined to a $K$--dimensional subspace of $\Cbb^M$. Thus, standard narrowband techniques used in array processing for detection may be used, see e.g. \cite{wax1985detection}.
\end{remark}

\section{A new consistent test statistic}
\label{section:test}

As we have seen in Theorem \ref{theorem:C_hat_y} and the related comments, the SCM $\hat{\C}_{\y}(\nu)$ behaves in the high-dimensional regime as a Wishart matrix with scale 
$\Xibs(\nu) = \S_{\v}(\nu)^{-\frac{1}{2}} \H(\nu)\H(\nu)^*\S_{\v}(\nu)^{-\frac{1}{2}} + \I_M$ being a fixed rank $K$ perturbation of the identity matrix. The behaviour of the eigenvalues for each $\nu$ of such matrix model is well-known since \cite{Baik2006} (and other related works such as the well-known BBP-phase-transition \cite{baik2005phase} or \cite{benaych2011eigenvalues}), and the rest of this section is devoted to the application of the results from \cite{Baik2006} in our frequency-domain detection context. A crucial point is to choose the particular frequency at which the above mentioned results will be used in order to obtain information on the behaviour of $\max_{\nu \in \Vcal_N} \lambda_1\left(\hat{\C}_{\y}(\nu)\right)$. For this, we have first to define some notations. We consider the fundamental function $\phi$ which already appears in \cite{Baik2006}:
\begin{align}
  \phi(x) =
  \begin{cases}
    \frac{(x+1)(x+c)}{x} & \quad \text{ if } x > \sqrt{c}
    \\
    \lambda^+ & \quad \text{ if } x \leq \sqrt{c}
  \end{cases}
                                \notag
\end{align}
where we recall that $\lambda^+ = (1+\sqrt{c})^2$ (see Corollary \ref{corollary:lss}). We notice that for all $x > \sqrt{c}$, $\phi(x) > \phi(\sqrt{c}) = \lambda^+$. Define as $\gamma(\nu)$ the maximum eigenvalue of the finite rank perturbation for each $\nu$, that is
\begin{align}
  \gamma(\nu) = \lambda_1\left(\S_{\v}(\nu)^{-\frac{1}{2}}\H(\nu)\H(\nu)^*\S_{\v}(\nu)^{-\frac{1}{2}}\right)
  \label{eq:gamma}
\end{align}
and let $\nu_N^* \in \Vcal_N$ such that
\begin{align}
  \nu_N^* \in \argmax_{\nu \in \Vcal_N} \gamma(\nu).
  \notag
\end{align}
We remark that $\gamma(\nu^*_N)$ may be interpreted as a certain SNR metric in the frequency domain. In the following, we study the behaviour of the largest eigenvalue of $\hat{\C}_{\y}(\nu^*_N)$, which requires the following additional assumption on $\gamma(\nu^*_N)$.
\begin{assumption}
  \label{assumption:spike}
  There exists $\gamma_{\infty} \geq 0$ such that
  \begin{align}
    \gamma(\nu^*_N)
    \xrightarrow[M\to\infty]{} \gamma_{\infty}.
    \notag
  \end{align}
\end{assumption}

Theorem \ref{theorem:C_hat_y} implies that the eigenvalues of $\hat{\C}_{\y}(\nu^*_N)$ have the same asymptotic behaviour as the corresponding eigenvalues of matrix $ \Xibs(\nu^*_N)^{\frac{1}{2}} \frac{\X(\nu^*_N)\X(\nu^*_N)^*}{B+1}\Xibs(\nu^*_N)^{\frac{1}{2}}$. Under Assumption \ref{assumption:spike}, \cite{baik2005phase}, \cite{Baik2006} or \cite{benaych2011eigenvalues} immediately imply the following result.
Note that since $\nu^*_N$ is unknown in practice,
this proposition is an intermediate theoretical result that will justify the detection test statistic introduced below.
\begin{proposition}
\label{proposition:spike}
  Under Assumptions \ref{assumption:rm}, \ref{assumption:sm}, \ref{assumption:H}, \ref{assumption:regime} and \ref{assumption:spike}, we have
  \begin{align}
    \lambda_1\left(\hat{\C}_{\y}(\nu_N^*)\right) \xrightarrow[M\to\infty]{a.s.} \phi(\gamma_{\infty})
\label{eq:spikelimit1}
  \end{align}
while 
\begin{align}
  \lambda_ {K+1}\left(\hat{\C}_{\y}(\nu_N^*)\right) \xrightarrow[M\to\infty]{a.s.} \lambda^+
\label{eq:spikelimit2}
\end{align}
and
\begin{align}
  \lambda_{M}\left(\hat{\C}_{\y}(\nu_N^*)\right) \xrightarrow[M\to\infty]{a.s.}  \lambda^-.
\label{eq:spikelimit3}
\end{align}
Moreover, if $\gamma_{\infty} = 0$, 
\begin{align}
  \limsup_{M\to\infty} \max_{\nu \in \Vcal_N}\lambda_1\left(\hat{\C}_{\y}(\nu)\right) \leq \lambda^+ \quad \text{a.s.}
  \label{eq:spikelimit4}
\end{align}

\end{proposition}

\begin{IEEEproof}
 It just remains to establish (\ref{eq:spikelimit4}), see Appendix \ref{appendix:proof_proposition_spike}.
\end{IEEEproof}
Since neither the intrinsic dimensionality $K$ of the useful signal $(\u_n)_{n \in \Zbb}$ nor the frequency $\nu^*_N$ are known in practice, we use the largest eigenvalue of the SCM maximized over all Fourier frequencies as a test statistic. This leads to the test statistic $T_\epsilon$ defined previously in \eqref{eq:LE} which we recall here:
\begin{align*}
  T_{\epsilon} = \mathbb{1}_{[\lambda^+ + \epsilon,\infty)}\left(\max_{\nu \in \Vcal_N} \lambda_1\left(\hat{\C}_{\y}(\nu)\right)\right).
\end{align*}
It turns out that this test statistics is consistent in the high-dimensional regime, as stated in the following result.
\begin{proposition}
\label{prop:test-consistency}
  Under Assumptions \ref{assumption:rm}, \ref{assumption:sm}, \ref{assumption:H}, \ref{assumption:regime} and \ref{assumption:spike}, and if
  under Hypothesis $\Hcal_1$,
  \begin{align}
    \gamma_{\infty} > \sqrt{c}
    \notag
  \end{align}
  then for all $0 < \epsilon < \phi(\gamma_{\infty})-\lambda^+$ and $i \in \{0,1\}$,
  \begin{align}
    \Pbb_i\left(\lim_{M\to\infty} T_{\epsilon} = i\right) = 1
    \notag
  \end{align}
  where $\Pbb_i$ is the underlying probability measure under Hypothesis $\Hcal_i$.
\end{proposition}
\begin{IEEEproof}
  Under Hypothesis $\Hcal_0$, since $\gamma_{\infty} = 0$, we directly apply \eqref{eq:spikelimit4} in Proposition \ref{proposition:spike} to obtain that for all $\epsilon > 0$,
  $T_{\epsilon} = 0$ with probability one, for all large $M$.
  Under Hypothesis $\Hcal_1$, we get
  \begin{align}
    \liminf_{M\to\infty} \max_{\nu \in \Vcal_N} \lambda_1\left(\hat{\C}_{\y}(\nu)\right)
    \geq
    \lim_{M\to\infty} \lambda_1\left(\hat{\C}_{\y}(\nu^*_N)\right)
    = \phi(\gamma_{\infty})
    \notag
  \end{align}
  with probability one. Since by assumption, $\phi(\gamma_{\infty}) > \lambda^+ + \epsilon$, we deduce that  $T_{\epsilon} = 1$ with probability one for all large $M$.
\end{IEEEproof}

\section{Simulations}
\label{section:simulations}

In this section, we provide some numerical illustrations of the approximation results of Section \ref{section:scm}. We will consider the case where the rank $K$ of the signal is equal to one and then the case where $K$ is strictly greater than one.

\subsection{Case $K=1$}

\begin{figure}[!h]
\includegraphics[width=\columnwidth]{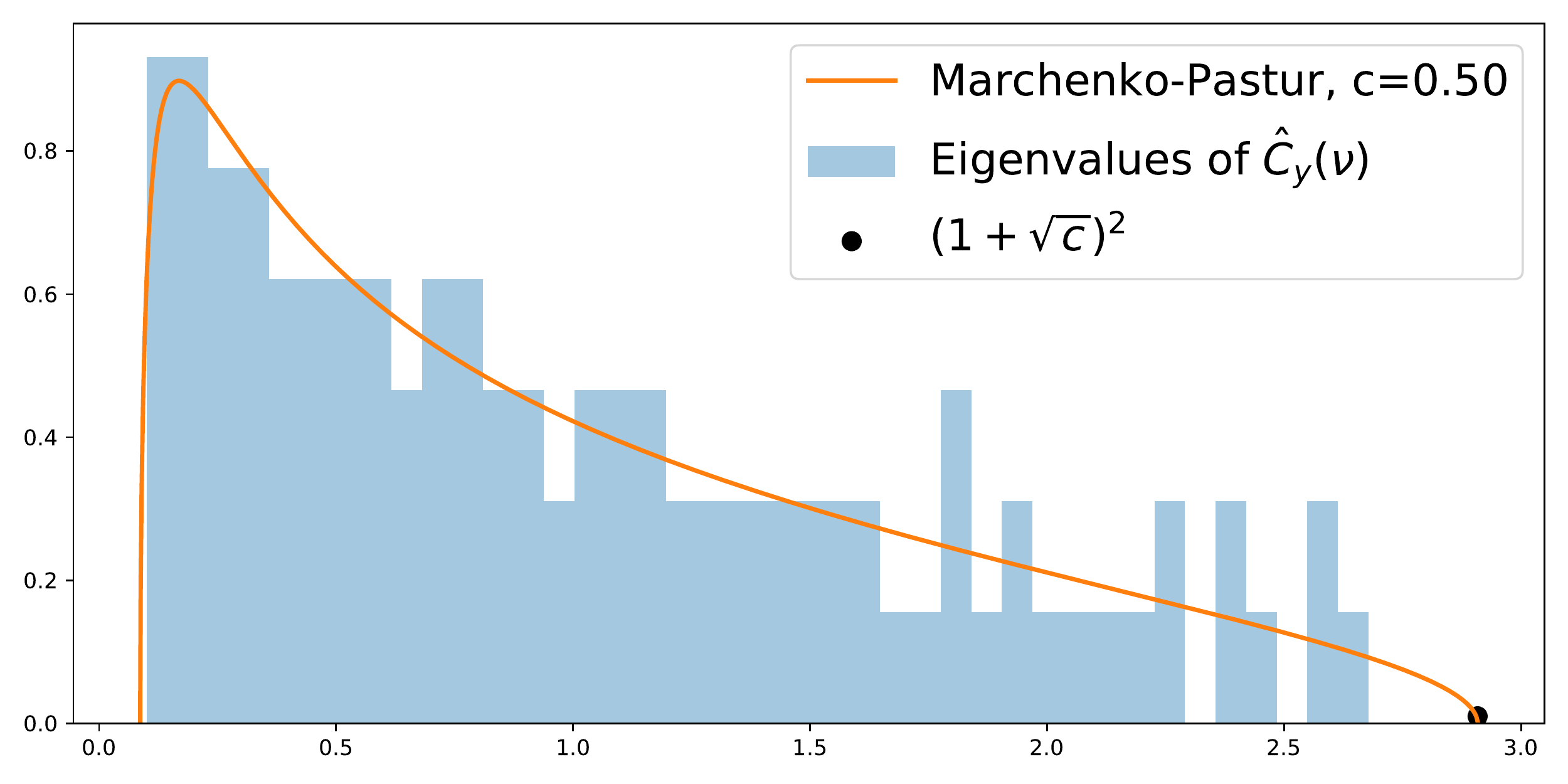}
\caption{Eigenvalue distribution of $\hat{\C}_{\y}(0)$ vs the density of the Marcenko-Pastur distribution with parameter $c=1/2$.} 
\label{figure:MP1} 
\end{figure}

\begin{figure}[!h]
\includegraphics[width=\columnwidth]{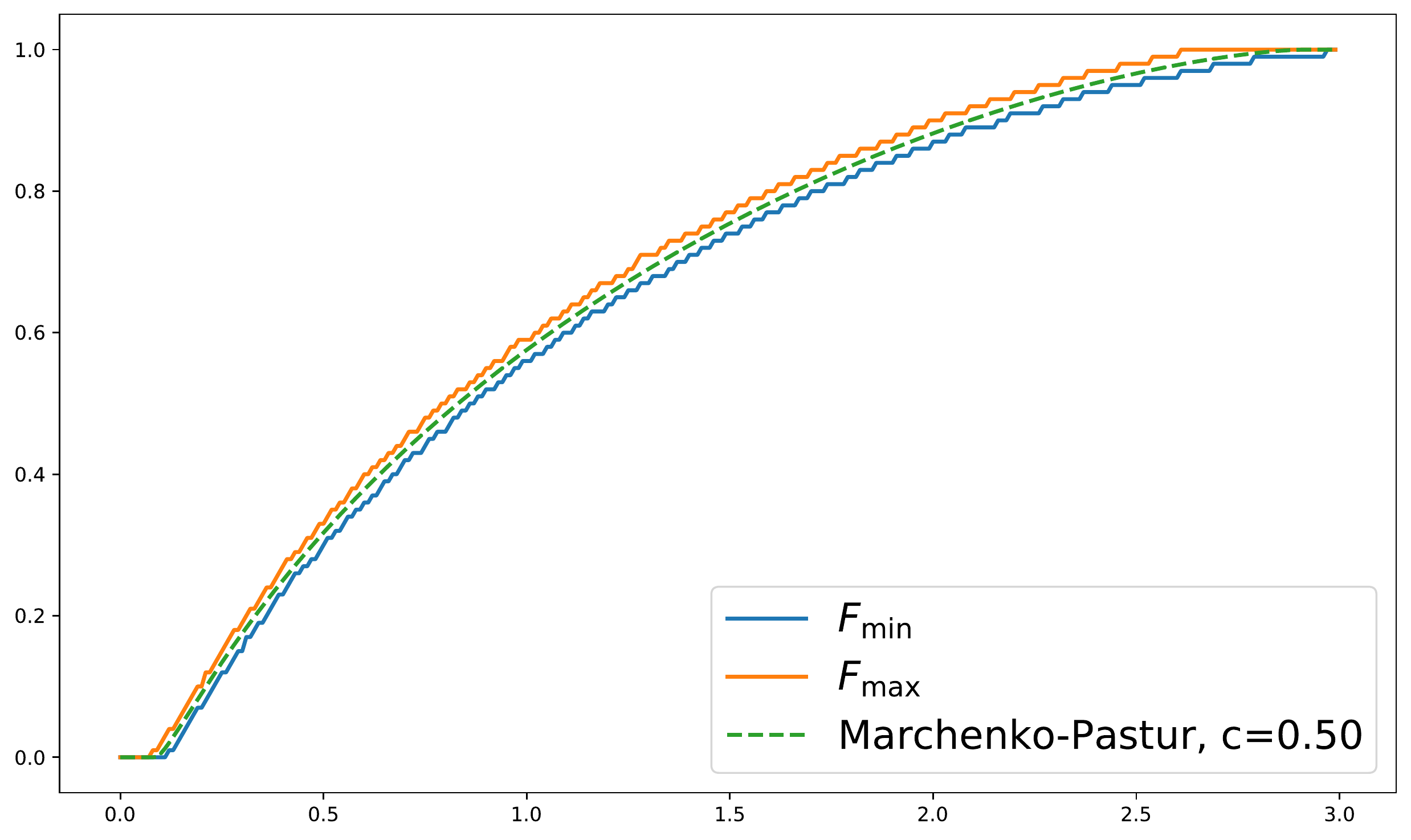}
\caption{Uniform convergence of the eigenvalue distribution of $\hat{\C}_{\y}(\nu)$ over $\nu\in\Vcal_N$ toward the Marcenko-Pastur distribution with parameter $c=1/2$.} 
\label{figure:MP2} 
\end{figure}

As in the numerical simulation presented in Section \ref{section:test_statistics_introduction}, each component of the noise $\v_n$ is generated as a Gaussian AR(1) process with $\theta=0.5$. The expression of its spectral density $s_m$ for all $m=1,\ldots,M$ is still given in \eqref{equation:s_m_simulations}. The useful signal is generated as an AR(1) process with $K=1$, $\H_k$ defined by \eqref{equation:H_k_simulations} and $\beta = \frac{10}{11}$. $C$ is again a positive constant used to tune the SNR. Note that, in this context, the SNR $\gamma(\nu)$ at frequency $\nu$ defined in \eqref{eq:gamma} takes the form
\begin{align}
  \gamma(\nu) = C \left|\frac{1-\theta\erm^{-\irm 2 \pi \nu}}{1-\beta\erm^{-\irm 2 \pi \nu}}\right|^2.
  \notag
\end{align}
Figures \ref{figure:MP1} and \ref{figure:MP2} illustrate the signal-free case $C=0$, and where $(N,M,B)= (20000, 100, 200)$.
In Figure \ref{figure:MP1}, we plot the histogram of the eigenvalues of $\hat{\C}_{\y}(\nu)$ for $\nu=0$. 
As predicted by Corollary \ref{corollary:lss} in the signal-free case, the empirical eigenvalue distribution of $\hat{\C}_{\y}(\nu)$ is well approximated by the Marcenko-Pastur distribution with shape parameter $c = 0.5 \approx M/(B+1)$.
Figure \ref{figure:MP2} further illustrates this convergence, where the cumulative distribution function (cdf) of the Marcenko-Pastur distribution is plotted against the two following quantities:
\begin{align*}
    F_{\min}(t) &= \min_{\nu\in\Vcal_N} \frac{1}{M}\sum_{\lambda_i(\hat{\C}(\nu))<t} \delta_{\lambda_i(\hat{\C}(\nu))} \\ 
    F_{\max}(t) &= \max_{\nu\in\Vcal_N} \frac{1}{M}\sum_{\lambda_i(\hat{\C}(\nu))<t} \delta_{\lambda_i(\hat{\C}(\nu))}.
\end{align*}
These two functions represent the maximum deviations (from above and below) over the frequencies $\nu\in\Vcal_N$ of the empirical spectral distribution of $\hat{\C}(\nu)$ against the Marcenko-Pastur distribution. As suggested by the uniform convergence in the frequency domain in Corollary \ref{corollary:lss}, the Marcenko-Pastur approximation in the high-dimensional regime is reliable over the whole set of Fourier frequencies. Note that the statement of Corollary \ref{corollary:lss} does not exactly match the setting used in Figure \ref{figure:MP2}, as the test function used here is not in $\Ccal^1((0,+\infty))$. 

\begin{figure}[!h]
\includegraphics[width=\columnwidth]{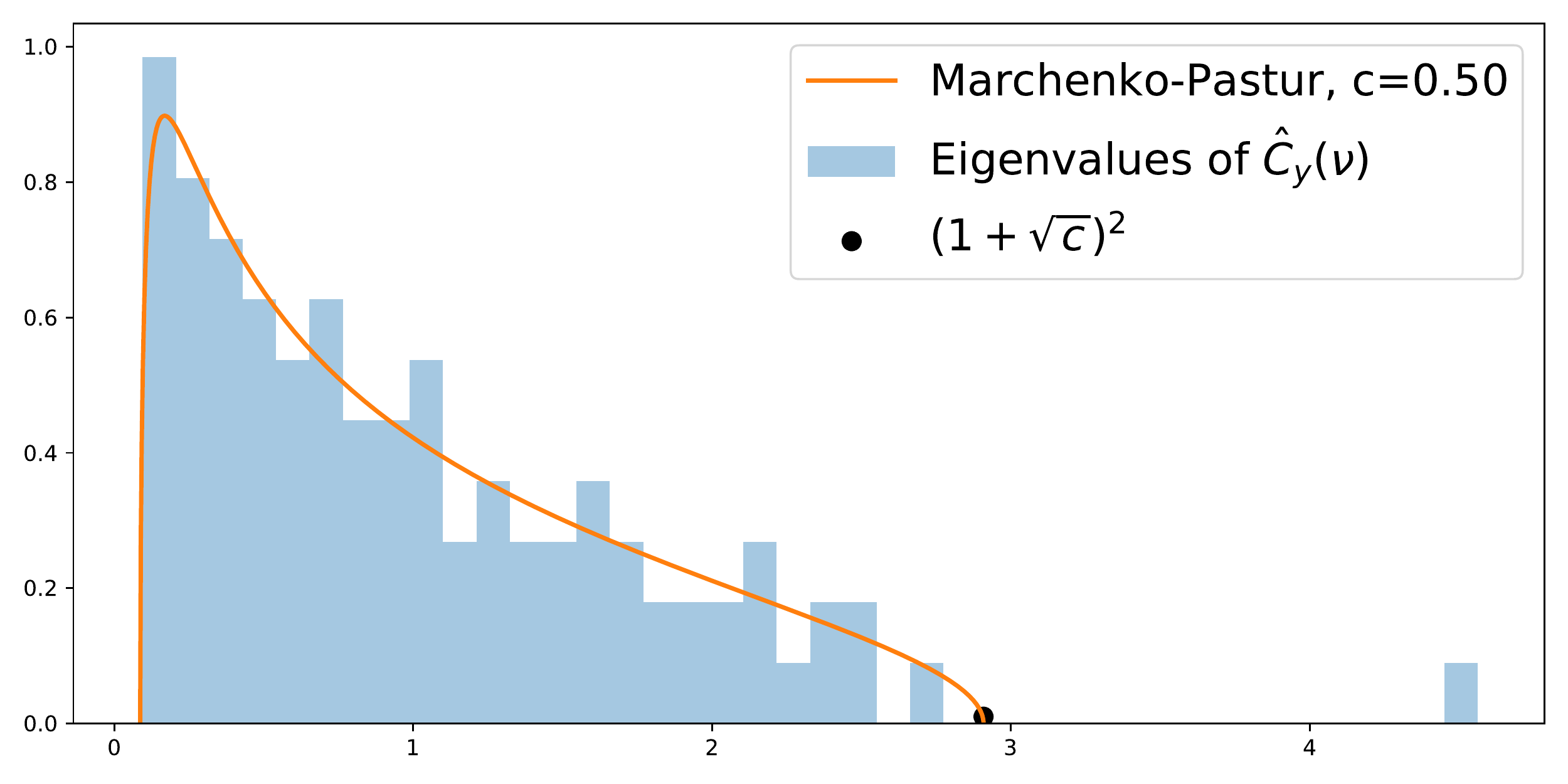}
\caption{Eigenvalue distribution of $\hat{\C}(\nu)$ vs Marcenko-Pastur distribution with parameter $c=1/2$ in the signal case.} 
\label{figure:signal_spike_vs_MP_1} 
\end{figure}

To illustrate the signal-plus-noise case and the results of Corollary \ref{corollary:lss} and Proposition \ref{proposition:spike}, we plot in Figure \ref{figure:signal_spike_vs_MP_1}, the histogram of the eigenvalues of $\hat{\C}_{\y}(\nu)$ for $\nu=0$, with $\gamma(0) = 2.9$. We see that the largest eigenvalue deviates from the right edge $(1+\sqrt{c})^2$ and is located around the value
$\phi\left(\gamma(0)\right) = 4.5$, as predicted by Proposition \ref{proposition:spike}, while all the other eigenvalues spread as the Marcenko-Pastur distribution, as predicted by
Corollary \ref{corollary:lss}. 

In order to compare the test statistic \eqref{eq:LE} with other frequency domain methods based on the SCM, we consider:
\begin{itemize}
\item the new test statistic \eqref{eq:LE}, denoted as LE (for largest eigenvalue),
\item two tests based on LSS of the SCM given by
  \begin{align}
    T'_{\epsilon} = \mathbb{1}_{[\epsilon,+\infty)}\left(\max_{\nu \in \Vcal_N} \left|L_{\varphi}(\nu)-\int_{\Rbb}\varphi(\lambda)f(\lambda)\drm\lambda\right|\right)
    \notag
  \end{align}
  where $L_{\varphi}$ and density $f$ are defined in \eqref{eq:lss} and Corollary \ref{corollary:lss} respectively, and with $\varphi(x) = (x-1)^2$ for the Frobenius norm test (denoted as LSS Frob.) and $\varphi(x) = \log(x)$ for the logdet test  (denoted as LSS logdet),
\item a test statistic based on the largest off-diagonal entry of the SCM:
  \begin{align}
    T^{''}_{\epsilon} = \mathbb{1}_{[\epsilon,+\infty)}\left(\max_{\nu \in \Vcal_N} \max_{\substack{i,j=1,\ldots,M \\ i < j}}\left|[\hat{\C}_{\y}(\nu)]_{i,j}\right|\right)
    \notag
  \end{align}
  denoted as MCC (for Maximum of Cross Coherence),
\end{itemize}
and where $\epsilon>0$ is some threshold. In Table \ref{table:power}, we provide, via Monte-Carlo simulations ($10000$ draws), the power of each of the four tests, calibrated so that the empirical type I error is equal to $0.05$. The results are provided for various values of $(N,M,B)$ chosen so that $M\in\{20,40,\ldots,180\}$, $N=M^2$ and $B=2M$. We set the SNR in the frequency domain as $\max_{\nu \in \Vcal_N} \gamma(\nu) = 2\sqrt{\frac{M}{B}} = 1.41$.


The LE test presents the best detection performance among the four candidates, whereas the MCC test does not seem to be adapted to the detection of this alternative. 
While it is proved in Corollary \ref{corollary:lss} that the test statistics based on the LSS of $\hat{\C}(\nu)$ can not asymptotically distinguish between $\Hcal_0$ and $\Hcal_1$, they remain sensible to a large variation of a single eigenvalue for finite values of $M$.
Consider for instance the Frobenius LSS test, where the test statistic is based on :
\[
   \max_{\nu \in \Vcal_N}  \left|\frac{1}{M}\sum_{m=1}^M (\lambda_m(\hat{\C}(\nu))-1)^2 - \int (\lambda-1)^{2} f(\lambda)d\lambda\right|
\]
where an explicit computation shows that $\int (\lambda-1)^{2} f(\lambda)d\lambda=c$. An $\Ocal(1)$ variation of $\lambda_1(\hat{\C}(\nu))$, the largest eigenvalue of $\hat{\C}(\nu)$, will lead to a variation of order $\Ocal(\frac{1}{M})$  of the above term. Therefore, the power of a LSS based test asymptotically converge towards zero, while having non-zero power for finite values of $M$ as it is visible on the results of Table \ref{table:power}.

\begin{table}
\caption{Power comparison, K=1, $\gamma(\nu_{N}^{*})=2\sqrt{\frac{1}{2}}$, type I error = 5\% \label{table:power}}
\centering
\begin{tabular}{ccccccc}
\toprule
      &     &  & LSS Frob. & LSS logdet &   MCC &  LE \\
N & M & B &               &            &       &       \\
\midrule
400   & 20  & 40  &           0.09 &        0.07 &  0.06 &  0.15 \\
1600  & 40  & 80  &           0.15 &        0.08 &  0.06 &  0.37 \\
3600  & 60  & 120 &           0.19 &        0.08 &  0.06 &  0.68 \\
6400  & 80  & 160 &           0.25 &        0.08 &  0.06 &  0.87 \\
10000 & 100 & 200 &           0.26 &        0.07 &  0.06 &  0.96 \\
14400 & 120 & 240 &           0.25 &        0.06 &  0.06 &  0.99 \\
19600 & 140 & 280 &           0.28 &        0.06 &  0.06 &  1.00 \\
25600 & 160 & 320 &           0.30 &        0.06 &  0.06 &  1.00 \\
32400 & 180 & 360 &           0.31 &        0.06 &  0.06 &  1.00 \\
\bottomrule
\end{tabular}
\end{table}

\subsection{Case $K>1$}
We eventually consider a model which have the flexibility to consider a signal with an arbitrary value of $K\ge1$.  We assume that matrices $(\H_l)_{l \geq 0}$ verify 
$\H_l = 0$ if $l > L$ for a certain integer $L$, and that the sequence of $M\times K$ matrices $(\H_l)_{0 \leq l\le L}$ is defined by:
\[
    \H_l = (C_1\w_{l,1},\ldots,C_K\w_{l,K})
\]
where the vectors $((\w_{l,k})_{l=0, \ldots, L})_{k=1, \ldots, K}$ are generated as independent realisations of $M$--dimensional vectors uniformly distributed on the unit sphere of $\mathbb{C}^{M}$ and where the $C_1 \geq C_2 \geq \ldots \geq C_K$ are positive constants used to tune 
the SNR of each of the $K$ sources at the desired level. 
Moreover, as 
the $K$ columns of each matrix $\H_l$ coincide with the realisations of mutually independent random vectors, the columns of $\H(\nu)$ are easily seen to be nearly orthogonal and to nearly share the same norm for each $\nu$ if $M$ is large enough. More precisely, for each $\nu$, it holds that $\H(\nu)^{*} \H(\nu) \rightarrow (L+1) \, \mathrm{Diag}(C_1, \ldots, C_K)$ when $M \rightarrow +\infty$. As the spectral densities of the components of the noise all coincide with $s(\nu) = \frac{1}{|1 - \theta e^{-2 i \pi \nu}|^{2}}$, the non-zero eigenvalues of $\S_{\v}(\nu)^{-\frac{1}{2}}\H(\nu)\H(\nu)^*\S_{\v}(\nu)^{-\frac{1}{2}}$ converge towards the $\left( (L+1) C_k/s(\nu)\right)_{k=1, \ldots, K}$ when $M$ increases. Therefore, the signal obtained by this model satisfies Assumption \ref{assumption:H}. Rather than just providing the performance of the test $T_{\epsilon}$ based on the maximum of the largest eigenvalue of $\hat{\C}_{\y}(\nu)$ proposed in this paper, we compare in the following $T_{\epsilon}$ with $T_{K,\epsilon}$ defined by 
$$
T_{K,\epsilon} = \mathbb{1}_{[K \lambda^+ +  \epsilon,\infty)}\left(\max_{\nu \in \Vcal_N} \sum_{k=1}^K\lambda_k\left(\hat{\C}_{\y}(\nu)\right)\right)
$$
which depends on the $K$ largest eigenvalues of $\hat{\C}_{\y}(\nu)$ rather than on the largest one. It is easy to 
generalize Proposition \ref{proposition:spike} and Proposition \ref{prop:test-consistency} 
in order to study the asymptotic properties of $T_{K,\epsilon}$. More precisely, for each $k=1, \ldots, K$, we define $\gamma_k(\nu)$ by
\begin{align}
\label{eq:definition_gamma_K}
  \gamma_k(\nu) = \lambda_k\left(\S_{\v}(\nu)^{-\frac{1}{2}}\H(\nu)\H(\nu)^*\S_{\v}(\nu)^{-\frac{1}{2}}\right)
\end{align}
and denote $\nu_{K,N}^*$ one of the frequency such that $\max_{\nu \in \Vcal_N} \sum_{k=1}^{K} \gamma_k(\nu) =  \sum_{k=1}^{K} \gamma_k(\nu_{K,N}^*)$. $\gamma_k(\nu)$ can of course be seen as a generalization of $\gamma(\nu)$ defined by (\ref{eq:gamma}). Then, under the extra assumption that for $k=1, \ldots, K$, $\gamma_k(\nu_{K,N}^*)$ converges towards a finite limit $\gamma_{k,\infty}$ (a condition which holds in the context of the present experiment because it is easily seen that $\gamma_k(\nu_{K,N}^{*}) \rightarrow (L+1) \, (1+\theta)^{2} C_k$), $\lambda_k\left(\hat{\C}_{\y}(\nu_{K,N}^*)\right)$ converges towards $\lambda^{+}$ if $\gamma_{k,\infty} \leq \sqrt{c}$ and towards $\phi(\gamma_{k,\infty}) > \lambda^{+}$ if $\gamma_{k,\infty} > \sqrt{c}$. It is easy to check that if $\gamma_{\infty} = \gamma_{1,\infty} > \sqrt{c}$, then the statistics $T_{K,\epsilon}$ also leads to a consistent test provided $0 <  \epsilon < \phi(\gamma_{\infty}) - \lambda_+$. While in practice the number of sources $K$ is unknown, it is interesting to evaluate the performance provided by $T_{K,\epsilon}$ which can be considered as an ideal reference. Intuitively, $T_{K,\epsilon}$ could lead to a better performance than $T_{\epsilon}$ when $\gamma_{k,\infty} > \sqrt{c}$ for $k=1, \ldots, K$, because, in this context, if $\hat{\nu}_{K,N}^*$ is a frequency that maximises $\sum_{k=1}^{K} \lambda_k\left( \hat{\C}_{\y}(\nu)\right)$, then $\liminf_{M \rightarrow +\infty} \lambda_k\left( \hat{\C}_{\y}(\nu_{K,N}^*)\right) > \lambda^{+}$. Therefore, the $K$ largest eigenvalues of $\hat{\C}_{\y}(\hat{\nu}_{K,N}^*)$ bring useful information to the detection of the useful signal. \\

In order to evaluate numerically the compared performance of $T_{\epsilon}$ and $T_{K,\epsilon}$ when $K$ is known, we first consider the case $K=2$, $L = 3$, and where $(\gamma_{1} + \gamma_{2})(\nu_{2,N}^{*}) = 3 \sqrt{c}$. Concerning the value of $(C_1,C_2)$, we consider the two following cases: $\frac{C_1}{C_2} = 1$ and $\frac{C_1}{C_2} = 4$. This corresponds respectively to the case where both sources contributes exactly the same on each sensor, and where the first source contributes much more than the second one. Tables \ref{table:power_K_5}, \ref{table:power_K_4} report the power of the proposed test (LE(1) represents $T_{\epsilon}$ and LE(2) represents $T_{2,\epsilon}$) against the LSS tests and the MCC test, with a type I error fixed at 5\%. When $\frac{C_1}{C_2}=4$, it can be expected that the most powerful source is dominant, and that $\gamma_{2}(\nu_{2,N}^{*}) < \sqrt{c}$. Therefore,  $\lambda_2\left(\hat{\C}_{\y}(\nu)\right)$ is likely to stay close to $\lambda^{+}$ for each $\nu$, so that the use of $T_{2,\epsilon}$ should not bring any extra performance. This intuition is confirmed by Table \ref{table:power_K_5}. When  $\frac{C_1}{C_2}=1$, $\gamma_{1}(\nu_{2,N}^{*})$ and $\gamma_{2}(\nu_{2,N}^{*})$ should be both close to $\frac{3}{2} \sqrt{c}$, thus suggesting that the two largest eigenvalues of $\hat{\C}_{\y}$ at the maximizing frequency $\hat{\nu}_{2,N}^{*}$ should also nearly coincide, and should escape from $[\lambda^{-}, \lambda^{+}]$. While the second eigenvalue brings here some information,  Table \ref{table:power_K_4} tends to indicate that $T_{\epsilon}$ has better performance than $T_{2,\epsilon}$. In the next experiment,  $(\gamma_{1} + \gamma_{2})(\nu_{2,N}^{*}) = 2 \sqrt{c}$. For $\frac{C_1}{C_2}=4$, the largest eigenvalue of $\hat{\C}_{\y}(\nu)$ is likely to be still dominant for each $\nu$, and Table \ref{table:power_K_3} confirms the better performance of $T_{\epsilon}$. When $\frac{C_1}{C_2}=1$, $\gamma_{1}(\nu_{2,N}^{*})$ and $\gamma_{2}(\nu_{2,N}^{*})$ should be both close to the detectability threshold $\sqrt{c}$, and Table \ref{table:power_K_2} this time shows that the use of $T_{2,\epsilon}$ leads to some improvement. For comparison, we also report the results of $T_{\epsilon}$ for $C_2 = 0$ in Table  \ref{table:power_K_1}.  


\begin{table}
\caption{Power comparison, $\frac{C_1}{C_2}=4$, $(\gamma_1+\gamma_{2})(\nu_{2,N}^{*})=3\sqrt{\frac{1}{2}}$, type I error = 5\% \label{table:power_K_5}}
\centering
\begin{tabular}{cccccccc}
\toprule
      &     &  & LSS Fr. & LSS ld &   MCC &  LE(1) & LE(2) \\
N & M & B &               &            &       &   &    \\
\midrule
100  & 10 & 20  &           0.31 &        0.18 &  0.16 &    0.42 &    0.37 \\
400  & 20 & 40  &           0.79 &        0.39 &  0.45 &    0.94 &    0.89 \\
900  & 30 & 60  &           0.94 &        0.49 &  0.53 &    1.00 &    0.99 \\
1600 & 40 & 80  &           0.98 &        0.50 &  0.55 &    1.00 &    1.00 \\
2500 & 50 & 100 &           0.99 &        0.52 &  0.55 &    1.00 &    1.00 \\
3600 & 60 & 120 &           1.00 &        0.51 &  0.43 &    1.00 &    1.00 \\
4900 & 70 & 140 &           1.00 &        0.55 &  0.37 &    1.00 &    1.00 \\
6400 & 80 & 160 &           1.00 &        0.54 &  0.28 &    1.00 &    1.00 \\
\bottomrule
\end{tabular}
\end{table}

\begin{table}
\caption{Power comparison, $\frac{C_1}{C_2}=1$, $(\gamma_1+\gamma_{2})(\nu_{2,N}^{*})=3\sqrt{\frac{1}{2}}$, type I error = 5\% \label{table:power_K_4}}
\centering
\begin{tabular}{cccccccc}
\toprule
      &     &  & LSS Fr. & LSS ld &   MCC &  LE(1) & LE(2) \\
N & M & B &               &            &       &   &    \\
\midrule
100  & 10 & 20  &           0.38 &        0.22 &  0.16 &    0.48 &    0.46 \\
400  & 20 & 40  &           0.58 &        0.30 &  0.30 &    0.75 &    0.73 \\
900  & 30 & 60  &           0.67 &        0.30 &  0.28 &    0.91 &    0.89 \\
1600 & 40 & 80  &           0.74 &        0.29 &  0.18 &    0.96 &    0.97 \\
2500 & 50 & 100 &           0.79 &        0.30 &  0.16 &    0.99 &    0.99 \\
3600 & 60 & 120 &           0.79 &        0.24 &  0.13 &    1.00 &    1.00 \\
4900 & 70 & 140 &           0.85 &        0.28 &  0.12 &    1.00 &    1.00 \\
\bottomrule
\end{tabular}
\end{table}

\begin{table}
\caption{Power comparison, $\frac{C_1}{C_2}=4$, $(\gamma_1+\gamma_{2})(\nu_{2,N}^{*})=2\sqrt{\frac{1}{2}}$, type I error = 5\% \label{table:power_K_3}}
\centering
\begin{tabular}{cccccccc}
\toprule
      &     &  & LSS Fr. & LSS ld &   MCC &  LE(1) & LE(2) \\
N & M & B &               &            &       &   &    \\
\midrule
100  & 10 & 20  &           0.15 &        0.10 &  0.10 &    0.21 &    0.20 \\
400  & 20 & 40  &           0.33 &        0.15 &  0.12 &    0.55 &    0.50 \\
900  & 30 & 60  &           0.39 &        0.15 &  0.17 &    0.75 &    0.71 \\
1600 & 40 & 80  &           0.52 &        0.16 &  0.14 &    0.94 &    0.90 \\
2500 & 50 & 100 &           0.54 &        0.15 &  0.14 &    0.98 &    0.97 \\
3600 & 60 & 120 &           0.56 &        0.13 &  0.13 &    1.00 &    0.99 \\
4900 & 70 & 140 &           0.55 &        0.13 &  0.10 &    1.00 &    1.00 \\
6400 & 80 & 160 &           0.62 &        0.11 &  0.10 &    1.00 &    1.00 \\
\bottomrule
\end{tabular}
\end{table}

\begin{table}
\caption{Power comparison, $\frac{C_1}{C_2}=1$, $(\gamma_1+\gamma_{2})(\nu_{2,N}^{*})=2\sqrt{\frac{1}{2}}$, type I error = 5\% \label{table:power_K_2}}
\centering
\begin{tabular}{cccccccc}
\toprule
      &     &  & LSS Fr. & LSS ld &   MCC &  LE(1) & LE(2) \\
N & M & B &               &            &       &      &  \\
\midrule
400   & 20  & 40  &           0.17 &        0.11 &  0.08 &    0.27 &    0.27 \\
1600  & 40  & 80  &           0.18 &        0.10 &  0.08 &    0.45 &    0.48 \\
3600  & 60  & 120 &           0.15 &        0.07 &  0.07 &    0.58 &    0.62 \\
6400  & 80  & 160 &           0.16 &        0.07 &  0.08 &    0.69 &    0.75 \\
10000 & 100 & 200 &           0.13 &        0.05 &  0.07 &    0.76 &    0.83 \\
14400 & 120 & 240 &           0.10 &        0.03 &  0.07 &    0.82 &    0.86 \\
19600 & 140 & 280 &           0.09 &        0.04 &  0.07 &    0.86 &    0.89 \\
25600 & 160 & 320 &           0.10 &        0.03 &  0.06 &    0.89 &    0.93 \\
32400 & 180 & 360 &           0.09 &        0.03 &  0.06 &    0.87 &    0.93 \\
\bottomrule
\end{tabular}
\end{table}

\begin{table}
\caption{Power comparison, $C_2=0$, $\gamma(\nu_{N}^{*})=2\sqrt{\frac{1}{2}}$, type I error = 5\% \label{table:power_K_1}}
\centering
\begin{tabular}{cccccccc}
\toprule
      &     &  & LSS Fr. & LSS ld &   MCC &  LE(1) & LE(2) \\
N & M & B &               &            &       &     &   \\
\midrule
100  & 10 & 20  &           0.19 &        0.12 &  0.12 &    0.26 &    0.22 \\
400  & 20 & 40  &           0.43 &        0.19 &  0.14 &    0.66 &    0.59 \\
900  & 30 & 60  &           0.51 &        0.19 &  0.19 &    0.88 &    0.83 \\
1600 & 40 & 80  &           0.62 &        0.20 &  0.15 &    0.97 &    0.95 \\
2500 & 50 & 100 &           0.65 &        0.18 &  0.17 &    0.99 &    0.99 \\
3600 & 60 & 120 &           0.68 &        0.16 &  0.12 &    1.00 &    1.00 \\
4900 & 70 & 140 &           0.71 &        0.16 &  0.13 &    1.00 &    1.00 \\
6400 & 80 & 160 &           0.75 &        0.17 &  0.12 &    1.00 &    1.00 \\
\bottomrule
\end{tabular}
\end{table}

This discussion tends to indicate that, even when $K > 1$ is assumed known, the use of the maximum over $\Vcal_N$ of the largest eigenvalue of $\hat{\C}_{\y}(\nu)$ does not introduce any significant loss of performance.

\section{Conclusion}

In this paper, we have studied the statistical behaviour of certain frequency-domain detection test statistics, based on the eigenvalues of a sample estimate of the SCM, in the high-dimensional regime in which both the dimension $M$ of the underlying signals and the number of samples $N$ converge to infinity at certain rates.
In particular, we have proved various approximation results showing that the sample SCM asymptotically behaves as a Wishart matrix. These results have been exploited to prove that test statistics based on LSS of the sample SCM are not consistent in the high-dimensional regime. A new test statistic relying on the largest eigenvalue of the sample SCM has also been proposed and proved to be consistent in the high-dimensional regime. Finally, numerical results have demonstrated that this new test statistic provides reasonable performance and outperforms other standard test statistics in situations where the dimension $M$ and the number of samples $N$ are large.

\bibliographystyle{IEEEtran}
\bibliography{references}

\begin{appendices}

\section{Useful results}
\label{appendix:useful_results}

In this section, we recall some useful results which will be constantly used in the proofs developed in the following sections.

The first result is based on a Chernoff bound for the $\chi^2$ distribution, and is also a special case of the well-known Hanson-Wright inequality describing the concentration of sub-Gaussian quadratic forms around their means (see \cite{Rudelson2013}).
\begin{lemma}
  \label{lemma:hanson-wright}
  Let $\z \sim \Ncal_{\Cbb^n}(\mathbf{0},\I_n)$ and $\Xibs$ a deterministic $n \times n$ complex matrix.
  Then there exists a constant $\kappa > 0$ independent of $n$ and $\Xibs$ such that for all $t \geq 0$,
  \begin{equation*}
    \Pbb\left(\left|\z^*\Xibs\z-\Ebb[\z^*\Xibs\z]\right| > t \right) \leq 2\ \exp\left(-\kappa\min\left\{\frac{t^2}{\|\Xibs\|_F^2},\frac{t}{\|\Xibs\|_2}\right\}\right).
  \end{equation*}
\end{lemma}
The second following result describes the behaviour of the largest and smallest eigenvalues of a standard Wishart matrix.
\begin{lemma}[{\cite[Proof of Lemma 7.3]{Haagerup2003}}]
  \label{lemma:haagerup}
  Let $\Z$ be a $M \times (B+1)$ matrix with i.i.d. $\Ncal_{\Cbb}\left(0,1\right)$ entries. Then under Assumption \ref{assumption:regime}, there exists a constant $C > 0$ independent of $M,B$
  such that for all $t > 0$,
  \begin{equation*}
    \Pbb\left(\lambda_1\left(\frac{\Z\Z^*}{B+1}\right) > \left(1 + \sqrt{\frac{M}{B+1}}\right)^2 + t\right)  \leq
      (B+1) \exp\left(- C (B+1) t^2\right)
  \end{equation*}
  and
  \begin{equation*}
    \Pbb\left(\lambda_M\left(\frac{\Z\Z^*}{B+1}\right) < \left(1 - \sqrt{\frac{M}{B+1}}\right)^2 - t\right) \leq
      (B+1) \exp\left(- C (B+1) t^2\right).
  \end{equation*}
\end{lemma}
We will mainly use Lemma \ref{lemma:haagerup} as follows; let $(\Z(\nu))_{\nu \in \Vcal_N}$ be a family of $M \times (B+1)$ random matrices such that
$\Z(\nu)$ has i.i.d. $\Ncal_{\Cbb}(0,1)$ (recall the definition of the index set $\Vcal_N$ in
\eqref{eq:V_set}), then from the union bound
\begin{align}
  &\Pbb\left(\max_{\nu \in \Vcal_N}\lambda_1\left(\frac{\Z(\nu)\Z(\nu)^*}{B+1}\right) > \left(1 + \sqrt{\frac{M}{B+1}}\right)^2 + t\right)
\notag\\
  & \leq
  \sum_{\nu \in \Vcal_N}
  \Pbb\left(\lambda_1\left(\frac{\Z(\nu)\Z(\nu)^*}{B+1}\right) > \left(1 + \sqrt{\frac{M}{B+1}}\right)^2 + t\right)
    \notag\\
  &\leq
    \exp\Bigl(-C(B+1)t^2 + \log(N(B+1))\Bigr).
    \notag
\end{align}
Using Assumption \ref{assumption:regime} and Borel-Cantelli lemma, we deduce that
\begin{align}
  \limsup_{M\to\infty} \max_{\nu \in \Vcal_N} \lambda_1\left(\frac{\Z(\nu)\Z(\nu)^*}{B+1}\right) \leq (1+\sqrt{c})^2
  \notag
\end{align}
with probability one.

\section{Proof of Theorem \ref{theorem:pure_signal}}
\label{appendix:proof_theorem_pure_signal}

\subsection{Reduction to $K=1$}

First, note that we may assume $K=1$ without loss of generality. Indeed, consider the decomposition
\begin{align}
	\u_n = \sum_{\ell = 1}^K \u_{n}^{(\ell)}
	\notag
\end{align}
where $\u_{n}^{(\ell)} = \sum_{k=0}^{+\infty} \h_{\ell,k} \epsilon_{n-k}^{(\ell)}$ and where $\h_{\ell,k}$ and $\epsilon_{n}^{(\ell)}$ are the $\ell$-th column of $\H_k$ and the $\ell$-th
entry of $\epsilonbs_n$ respectively.
Moreover, Assumption \ref{assumption:H} implies that
\begin{align}
	\sup_{M \geq 1} \sum_{k \in \Zbb} (1+|k|) \|\h_{\ell,k}\|_2 < \infty
	\notag
\end{align}
From the fact that $K$ is fixed with respect to $N$ (Assumption \ref{assumption:regime}) and
\begin{align*}
    \max_{\nu \in \Vcal_N} \left\|\Sigmabs_{\u}\left(\nu\right) - \H\left(\nu\right)\Sigmabs_{\epsilonbs}\left(\nu\right)\right\|_2
	\leq \sum_{\ell=1}^K
	\max_{\nu \in \Vcal_N}
	\left\|
		\Sigmabs_{\u^{(\ell)}}\left(\nu\right) 
		- \h_{\ell}\left(\nu\right) \Sigmabs_{\epsilon^{(\ell)}}\left(\nu\right)
	\right\|_2
\end{align*}
where $\Sigmabs_{\u^{(\ell)}}\left(\nu\right)$, $\h_{\ell}\left(\nu\right)$, $\Sigmabs_{\epsilon^{(\ell)}}\left(\nu\right)$ are defined as $\Sigmabs_{\u}\left(\nu\right)$, $\H\left(\nu\right)$, $\Sigmabs_{\epsilonbs}\left(\nu\right)$ respectively, Theorem \ref{theorem:pure_signal} is proved if we can show that
\begin{align}
  \max_{\nu \in \Vcal_N}
	\left\|
		\Sigmabs_{\u^{(\ell)}}\left(\nu\right) 
		- \h_{\ell}\left(\nu\right) \Sigmabs_{\epsilon^{(\ell)}}\left(\nu\right)
  \right\|_2
  \xrightarrow[M\to\infty]{a.s.} 0
  \notag
\end{align}
for all $\ell=1,\ldots,K$. Therefore, we assume for the remainder of the proof that 
\begin{align}
	\u_n = \sum_{k \in \Zbb} \h_k \epsilon_{n-k},
	\notag
\end{align}
where 
\begin{itemize}
	\item $(\h_k)_{k \in \Zbb}$ is a filter, with $\h_k \in \Cbb^M$ and such that 
	\begin{align}
		\sup_{M \geq 1} \sum_{k\in \Zbb} (1+|k|) \left\|\h_k\right\|_2 < \infty.
		\label{assumption:h}
	\end{align} 
	\item $(\epsilon_n)_{n \in \Zbb}$ is a scalar standard complex Gaussian white noise.
\end{itemize}

\subsection{Reduction to $B=1$}
Let $\h(\nu) =\sum_{k \in \Zbb} \h_k \erm^{-\irm 2 \pi \nu k}$ and $$\xi_{\epsilon}(\nu) = \frac{1}{\sqrt{N}}\sum_{n=0}^{N-1} \epsilon_n \erm^{-\irm 2 \pi \nu n}.$$
From \eqref{assumption:h} and Assumption \ref{assumption:regime}, a first-order Taylor expansion of $b \mapsto \h\left(\nu + \frac{b}{N}\right)$ at $0$ leads to
\begin{align}
	\sup_{\nu \in [0,1]} \max_{b \in  \{-\frac{B}{2},\ldots,\frac{B}{2}\}}
  \left\|\h(\nu) - \h\left(\nu + \frac{b}{N}\right)\right\|_2
  &= \Ocal\left(\frac{B}{N}\right)
    \notag = \Ocal\left(\frac{1}{N^{1-\alpha}}\right).
    \notag
\end{align}
Moreover, from Lemma \ref{lemma:hanson-wright} applied to the random vector
\begin{align}
  \z = \left(\xi_{\epsilon}\left(\nu - \frac{B}{2N}\right),\ldots,\xi_{\epsilon}\left(\nu + \frac{B}{2N}\right)\right)^T \sim \Ncal_{\Cbb^{B+1}}\left(\mathbf{0},\I_{B+1}\right)
  \notag
\end{align}
and matrix $\Xibs = \frac{\I_{B+1}}{B+1}$, there exists some constant $\kappa$ independent of $M$ such that for all $t \geq 2$,
\begin{align}
  \Pbb\left(\max_{\nu \in \Vcal_N}
    \frac{1}{B+1} \sum_{b=-B/2}^{B/2} \left|\xi_{\epsilon}\left(\nu + \frac{b}{N}\right)\right|^2 > t\right) \leq
	N \Pbb\left(\frac{1}{B+1} \sum_{b=-B/2}^{B/2} \left|\xi_{\epsilon}\left(\frac{b}{N}\right)\right|^2 > t\right) \leq
	N \exp\left(- \kappa B \right)
	\notag
\end{align}
and  Borel-Cantelli lemma together with Assumption \ref{assumption:regime} imply
\begin{align}
  \max_{\nu \in \Vcal_N} \frac{1}{B+1} \sum_{b=-B/2}^{B/2} \left|\xi_{\epsilon}\left(\nu + \frac{b}{N}\right)\right|^2 = \Ocal\left(1\right)
  \notag
\end{align}
with probability one.
Defining
\begin{align}
  \Sigmabs_{\epsilon}(\nu) =
  \frac{1}{\sqrt{B+1}} \left(\xi_{\epsilon}\left(\nu - \frac{B}{2N}\right),\ldots, \xi_{\epsilon}\left(\nu + \frac{B}{2N}\right)\right)
	\notag
\end{align}
as well as
\begin{align}
  \Phibs(\nu) =
  \frac{1}{\sqrt{B+1}}
  \left[
  \phibs\left(\nu - \frac{B}{2N}\right),
  \ldots, 
  \phibs\left(\nu + \frac{B}{2N}\right)
  \right]
  \notag
\end{align}
with $\phibs(\nu) = \h(\nu) \xi_{\epsilon}(\nu)$,  we therefore have the control
\begin{align}
  \max_{\nu \in \Vcal_N}
  \left\|
  \h(\nu)\Sigmabs_{\epsilon}(\nu) 
  - 
  \Phibs(\nu)
    \right\|_2 \leq
    &\sup_{\nu \in [0,1]}
    \max_{b \in  \{-\frac{B}{2},\ldots,\frac{B}{2}\}} \left\|\h(\nu) - \h\left(\nu + \frac{b}{N}\right)\right\|_2 \sqrt{\max_{\nu \in \Vcal_N} \frac{1}{B+1} \sum_{b=-B/2}^{B/2} \left|\xi_{\epsilon}\left(\nu + \frac{b}{N}\right)\right|^2}
    \notag\\
\notag  &= \Ocal\left(\frac{1}{N^{1-\alpha}}\right) \text{ a.s.}
    \\
    &\notag \xrightarrow[M\to\infty]{a.s.} 0.
  \notag
\end{align}
Finally, since the spectral norm of a matrix is bounded by its Frobenius norm,
\begin{align}
  \max_{\nu \in \Vcal_N}\left\|\Sigmabs_{\u}(\nu) - \Phibs(\nu)\right\|_2  &\le \sqrt{\frac{1}{B+1} \sum_{b=-B/2}^{B/2} \left\|\xibs_{\u}\left(\nu+\frac{b}{N}\right)-\phibs\left(\nu+\frac{b}{N}\right)\right\|_2^2}
    \notag\\
  &\leq
    \max_{\nu \in \Vcal_N} \left\|\xibs_{\u}\left(\nu\right) -\phibs\left(\nu\right)\right\|_2.
    \notag
\end{align}
Theorem 1 is proven if we show that 
\begin{align}
	\max_{\nu \in \Vcal_N} \left\|\xibs_{\u}\left(\nu\right) - \h\left(\nu\right)\xibs_{\epsilon}\left(\nu\right)\right\|_2	
	\xrightarrow[N\to\infty]{a.s.} 0.
	\notag
\end{align}

\subsection{Periodization}

For all integer $n$, let $[n]$ denotes the integer contained in $\{0,\ldots,N-1\}$ such that $[n] \equiv n \pmod N$  and define
\begin{align}
	\tilde{\u}_n = \sum_{k \in \Zbb} \h_k \epsilon_{[n-k]}
	\notag
\end{align}
where $(\tilde{\u}_n)_{n \in \Zbb}$ represents the circular convolution between $(\h_k)_{k \in \Zbb}$ and $(\epsilon_n)_{n \in \Zbb}$. 
If $\xibs_{\tilde{\u}}(\nu) = \frac{1}{\sqrt{N}}\sum_{n=0}^{N-1} \tilde{\u}_n \erm^{-\irm 2 \pi n \nu}$, then the equality
\begin{align}
	\xibs_{\tilde{\u}}(\nu) = \h(\nu) \xi_{\epsilon}(\nu)
	\notag
\end{align}
holds for all $\nu \in \Vcal_N$.
It is straightforward to check that
\begin{align}
  \xibs_{\tilde{\u}}(\nu)-\xibs_{\u}(\nu) = \deltabs(\nu) + \check{\deltabs}(\nu)
  \notag
\end{align}
where
\begin{align}
  \deltabs(\nu)
  &=
  \frac{1}{\sqrt{N}} \sum_{k=1}^{N-1} \h_k \sum_{p=1}^k \left(\epsilon_{[-p]} - \epsilon_{-p}\right) \erm^{-\irm 2 \pi \nu (k-p)}
  \notag\\
  &+
    \frac{1}{\sqrt{N}} \sum_{k=N}^{+\infty} \h_k \sum_{p=0}^{N-1} \left(\epsilon_{[p-k]} - \epsilon_{p-k}\right) \erm^{-\irm 2 \pi \nu p}
	\notag
\end{align}
and
\begin{align}
  \check{\deltabs}(\nu)
    =  &\frac{1}{\sqrt{N}} \sum_{k=1}^{N-1} \h_{-k} \sum_{p=1}^k \left(\epsilon_{[N+p-1]} - \epsilon_{N+p-1}\right) \erm^{-\irm 2 \pi \nu (N-1+p-k)}
  \notag\\
  &+
    \frac{1}{\sqrt{N}} \sum_{k=N}^{+\infty} \h_{-k} \sum_{p=0}^{N-1} \left(\epsilon_{[p+k]} - \epsilon_{p+k}\right) \erm^{-\irm 2 \pi \nu p}
	\notag
\end{align}
Theorem \ref{theorem:pure_signal} is proved if we can show that
\begin{align}
  \max_{\nu \in \Vcal_N} \left\|\deltabs(\nu)\right\|_2 \xrightarrow[M\to\infty]{a.s.} 0
  \label{eq:conv_delta}
\end{align}
and
\begin{align}
  \max_{\nu \in \Vcal_N} \left\|\check{\deltabs}(\nu)\right\|_2 \xrightarrow[M\to\infty]{a.s.} 0.
  \label{eq:conv_delta_check}
\end{align}
In the remainder, we only prove \eqref{eq:conv_delta} and omit the details for \eqref{eq:conv_delta_check} whose treatment is similar. To that end, we define
\begin{align}
  &\deltabs_1(\nu) = \frac{1}{\sqrt{N}} \sum_{k=1}^{N-1} \h_k \sum_{p=1}^k \left(\epsilon_{[-p]} - \epsilon_{-p}\right) \erm^{-\irm 2 \pi \nu (k-p)} 
  \notag \\
  &\deltabs_2(\nu) =  \frac{1}{\sqrt{N}} \sum_{k=N}^{+\infty} \h_k \sum_{p=0}^{N-1} \left(\epsilon_{[p-k]} - \epsilon_{p-k}\right) \erm^{-\irm 2 \pi \nu p}.
  \notag
\end{align}

\subsection{Control of $\deltabs_1(\nu)$}

For $p \in \{1,\ldots,N-1\}$, let 
\begin{align}
	z_p(\nu) = \left(\epsilon_{[-p]} - \epsilon_{-p}\right)\erm^{\irm 2 \pi \nu p} = \left(\epsilon_{N-p} - \epsilon_{-p}\right)\erm^{\irm 2 \pi \nu p}.
	\notag
\end{align}
Then $z_1(\nu),\ldots,z_{N-1}(\nu)$ are i.i.d. $\Ncal_{\Cbb}(0,2)$ and by rearranging the sums in $\deltabs_1(\nu)$, we have
\begin{align}
	\deltabs_1(\nu) = \sum_{p=1}^{N-1} z_p(\nu) \g_p(\nu)
	\notag
\end{align}
with 
\begin{align}
	\g_p(\nu) = \frac{1}{\sqrt{N}} \sum_{k=p}^{N-1} \h_k \erm^{-\irm 2 \pi k \nu}.
	\notag
\end{align}
Therefore, $\deltabs_1(\nu) \sim \Ncal_{\Cbb^M}\left(\mathbf{0},\G(\nu)\right)$ with
\begin{align}
  \G(\nu) = 2 \sum_{p=1}^{N-1}\g_p(\nu)\g_p(\nu)^*.
  \notag
\end{align}
Moreover,
\begin{align*}
	\Ebb \left\|\deltabs_1(\nu)\right\|_2^2 = \tr \G(\nu)
    \leq
          \frac{2}{N} \sum_{p=1}^{N-1} 
          \left(
          \sum_{k=p}^{N-1} \left\|\h_k\right\|_2^2 
          + 2 \sum_{p \leq k < k' \leq N-1} \left\|\h_k\right\|_2\left\|\h_{k'}\right\|_2
          \right)
\end{align*}
and a straightforward rearrangement together with \eqref{assumption:h} leads to
\begin{align}
	\max_{\nu \in [0,1]} \Ebb \left\|\deltabs_1(\nu)\right\|_2^2
   &\leq 
	\frac{2}{N} \sum_{k=1}^{N-1} k \left\|\h_k\right\|_2^2 + \frac{4}{N} \sum_{1 \leq k < k' \leq N-1} \sqrt{k}\sqrt{k'} \left\|\h_k\right\|_2\left\|\h_{k'}\right\|_2
   \notag\\
        &=
          \frac{2}{N}\left(\sum_{k=1}^{N-1} \sqrt{k} \left\|\h_k\right\|_2\right)^2
          \notag \\
          &= \Ocal\left(\frac{1}{N}\right).
	\notag
\end{align}
where we used that $k\le \sqrt{k}\sqrt{k'}$ for $k'\geq k$. Additionally,
\begin{align}
  \max_{\nu \in [0,1]} \left\|\G(\nu)\right\|_2 \leq \max_{\nu \in [0,1]} \Tr \G(\nu) = \Ocal\left(\frac{1}{N}\right)
  \notag
\end{align}
and
\begin{align}
  \max_{\nu \in [0,1]} \left\|\G(\nu)\right\|_F \leq \sqrt{M} \max_{\nu \in [0,1]} \left\|\G(\nu)\right\|_2 = \Ocal\left(\frac{\sqrt{M}}{N}\right).
  \notag
\end{align}
Using Lemma \ref{lemma:hanson-wright}, there exists a constant $\kappa > 0$ independent of $M, (\h_k)_{k \in \Zbb}$ such that for all $t > 0$,
\begin{align*}
	\Pbb
	\left(
		\max_{\nu \in \Vcal_N} 
		\left|\left\|\deltabs_1(\nu)\right\|_2^2 - \Ebb \left\|\deltabs_1(\nu)\right\|_2^2\right|
		> t
   \right) \leq 
	2 N \max_{\nu \in \Vcal_N} 
	\exp\left(- \kappa \min\left(\frac{t^2}{\left\|\G(\nu)\right\|_F^2}, \frac{t}{\left\|\G(\nu)\right\|_2}\right)\right).
\end{align*}
Applying Assumption \ref{assumption:regime} and Borel-Cantelli lemma, it follows that
\begin{align}
	\max_{\nu \in \Vcal_N} \left|\left\|\deltabs_1(\nu)\right\|_2^2 - \Ebb \left\|\deltabs_1(\nu)\right\|_2^2\right|
	\xrightarrow[N\to\infty]{a.s.} 0.
	\notag
\end{align}
Finally, we deduce that
\begin{align}
	\max_{\nu \in \Vcal_N}\left\|\deltabs_1(\nu)\right\|_2^2
	&
	\leq
	\max_{\nu \in \Vcal_N} \Ebb \left\|\deltabs_1(\nu)\right\|_2^2 
  + \max_{\nu \in \Vcal_N} \left|\left\|\deltabs_1(\nu)\right\|_2^2 - \Ebb \left\|\deltabs_1(\nu)\right\|_2^2\right|\xrightarrow[N\to\infty]{a.s.} 0.
	\notag
\end{align}

\subsection{Control of $\deltabs_2(\nu)$}

We first split $\deltabs_2(\nu)$ in the following two parts
\begin{align}
	\deltabs_2(\nu) = \deltabs_{2,1}(\nu) + \deltabs_{2,2}(\nu)
	\notag
\end{align}
where 
\begin{align}
	&\deltabs_{2,1}(\nu) = 
	\frac{1}{\sqrt{N}} \sum_{k=N}^{+\infty} \h_k \sum_{p=0}^{N-1} \epsilon_{[p-k]} \erm^{-\irm 2 \pi p \nu}
	\notag \\
	&\deltabs_{2,2}(\nu) = 
	\frac{1}{\sqrt{N}} \sum_{k=N}^{+\infty} \h_k \sum_{p=0}^{N-1} \epsilon_{p-k} \erm^{-\irm 2 \pi p \nu}.
	\notag
\end{align}
We remark that $\deltabs_{2,1}(\nu)$ only involves the $N$ i.i.d. random variables $\epsilon_0,\ldots,\epsilon_{N-1}$ and that
\begin{align}
  \deltabs_{2,1}(\nu) = \sum_{p=0}^{N-1} \epsilon_p \tilde{\g}_p(\nu)
  \notag
\end{align}
with $\tilde{\g}_p(\nu)$ defined as
\begin{align}
  \tilde{\g}_p(\nu) = \frac{1}{\sqrt{N}} \sum_{k=N}^{+\infty} \h_k \erm^{-\irm 2\pi \nu [p+k]}.
  \notag
\end{align}
It is clear that
\begin{align}
  \max_{p=1,\ldots,N}\max_{\nu \in [0,1]} \|\tilde{\g}_p(\nu)\|_2
  &\leq
    \frac{1}{\sqrt{N}} \sum_{k=N}^{+\infty} \left\|\h_k\right\|_2 \leq
    \frac{1}{N^{3/2}} \sum_{k=N}^{+\infty} k\left\|\h_k\right\|_2
    \notag
\end{align}
and from \eqref{assumption:h},
\begin{align}
  \max_{p=1,\ldots,N}\max_{\nu \in [0,1]} \|\tilde{\g}_p(\nu)\|_2
  = o\left(\frac{1}{N^{3/2}}\right)
  \notag
\end{align}
Thus $\deltabs_{2,1}(\nu) \sim \Ncal_{\Cbb^M}\left(\mathbf{0},\tilde{\G}(\nu)\right)$ with
$\tilde{\G}(\nu) = \sum_{p=0}^{N-1} \tilde{\g}_p(\nu) \tilde{\g}_p(\nu)^*$ and
\begin{align}
  \max_{\nu \in [0,1]} \tr \tilde{\G}(\nu) = o\left(\frac{1}{N^2}\right)
  \notag
\end{align}
as $M\to\infty$.
Using Lemma \ref{lemma:hanson-wright} as for the control of $\deltabs_1(\nu)$ in the previous section, we end up with
\begin{align}
  \max_{\nu \in \Vcal_N} \left\|\deltabs_{2,1}(\nu)\right\|_2 \xrightarrow[M\to\infty]{a.s.} 0.
  \notag
\end{align}
We now consider the term $\deltabs_{2,2}(\nu)$, which involves the sequence of random variables $(\epsilon_{-n})_{n \geq 1}$.
For all $k \geq N$, set
\begin{align}
	\chibs_k = \frac{1}{\sqrt{N}} \h_k \sum_{p=0}^{N-1} \epsilon_{p-k} \erm^{-\irm 2 \pi p \nu}
	\notag
\end{align}
and consider the sequence $(\chibs_p)_{k \geq N}$. Using Assumption \ref{assumption:H},
\begin{align*}
	\sum_{k=N}^{+\infty} \|\chibs_k\|_2 &\le 
   \sum_{k=N}^{+\infty}\sqrt{k}\|\h_k\|_2 \left|\frac{1}{N} \sum_{p=0}^{N-1} \epsilon_{p-k} \erm^{-\irm 2 \pi p \nu} \right|
	\\ 
	&\le \left(\sup_{k\ge N} \left|\frac{1}{N} \sum_{p=0}^{N-1} \epsilon_{p-k} \erm^{-\irm 2 \pi p \nu} \right|\right) \sum_{k=N}^{+\infty}\sqrt{k}\|\h_k\|_2 \\
	&< +\infty \text{ a.s.}
\end{align*}
since for any $k$, by the gaussianity of the $\epsilon_{p-k}$, $\sup_{\nu\in\Vcal_N}\left|\frac{1}{N} \sum_{p=0}^{N-1} \epsilon_{p-k} \erm^{-\irm 2 \pi p \nu} \right|$ converges almost surely towards $0$ as $N\to+\infty$ by the law of the large numbers, so it remains almost surely bounded for any finite $N$. This implies that the family $(\chibs_k)_{k \geq N}$ is a.s. absolutely summable. Therefore, we can rearrange the series defining $\deltabs_{2,2}(\nu)$ and write
\begin{align}
	\deltabs_{2,2}(\nu) = \sum_{p=1}^{+\infty} \epsilon_{-p} \check{\g}_p(\nu)
	\notag
\end{align}
with probability one, where this time $\g_p(\nu)$ is defined for all $p \geq 1$ as
\begin{align}
  &\check{\g}_p(\nu) =
    \notag\\
  &\quad
 	\begin{cases}
		\frac{1}{\sqrt{N}} \sum_{k=0}^{p-1} \h_{k+N} \ \erm^{-\irm 2 \pi (N+k-p)\nu} & \text{if } p \in \{1,\ldots,N\}
		\notag\\
		\frac{1}{\sqrt{N}} \sum_{k=0}^{N-1} \h_{p+k} \ \erm^{-\irm 2 \pi k\nu} & \text{if } p \geq N+1.
		\notag
 	\end{cases}.
	\notag
\end{align}
Again,
\begin{align}
  \sup_{p \geq 1}\max_{\nu \in [0,1]} \left\|\check{\g}_p(\nu)\right\|_2 = o\left(\frac{1}{N}\right) 
\end{align}
$\deltabs_{2,2}(\nu) \sim \Ncal_{\Cbb^M}\left(\mathbf{0}, \check{\G}(\nu)\right)$, where
\begin{align}
  \check{\G}(\nu) = \sum_{p=1}^{+\infty} \check{\g}_p(\nu)\check{\g}_p(\nu)^*
  \notag
\end{align}
and such that $\tr \check{\G}(\nu) = o\left(\frac{1}{N}\right)$.
Thus, using Lemma \ref{lemma:hanson-wright} also yields
\begin{align}
	\max_{\nu \in \Vcal_N} \left\|\deltabs_{2,2}(\nu)\right\|_2 \xrightarrow[M\to\infty]{a.s.} 0.
	\notag
\end{align}
This concludes the proof of Theorem \ref{theorem:pure_signal}.

\section{Proof of Theorem \ref{theorem:C_hat_y}}
\label{appendix:proof_theorem_SCM}

To prove Theorem \ref{theorem:C_hat_y}, we need as a preliminary step to study the behaviour of the renormalization by $\dg(\hat{\S}_{\y}(\nu))^{-\frac{1}{2}}$ in the SCM.
\begin{lemma}
   \label{lemma:diag_S_hat_y}
   Under Assumptions \ref{assumption:rm}, \ref{assumption:H} and \ref{assumption:regime}, we have
   \begin{align}
    \max_{\nu \in \Vcal_N} \left\|\dg\left(\hat{\S}_{\y}(\nu)\right) - \S_{\v}(\nu)\right\|_2
    \xrightarrow[M\to\infty]{a.s.} 0
     \label{eq:diag_S_hat_y}
   \end{align}
   as well as
   \begin{align}
    \max_{\nu \in \Vcal_N} \left\|\dg\left(\hat{\S}_{\y}(\nu)\right)^{-\frac{1}{2}} - \S_{\v}(\nu)^{-\frac{1}{2}}\right\|_2
    \xrightarrow[M\to\infty]{a.s.} 0
     \label{eq:diag_S_hat_y_sqrt}
   \end{align}
\end{lemma}
\begin{IEEEproof}
  To prove \eqref{eq:diag_S_hat_y}, we establish successively
    \begin{equation}
      \label{eq:inter-step1}
      \max_{\nu \in \Vcal_N} \left\|\dg\left(\hat{\S}_{\y}(\nu)\right) - \dg\left(\S_{\y}(\nu)\right)\right\|_2 \xrightarrow[M\to\infty]{a.s.} 0
    \end{equation}
    as well as 
    \begin{equation}
      \label{eq:inter-step2}
      \max_{\nu \in \Vcal_N} \left\|\dg\left(\S_{\y}(\nu)\right) -  \S_{\v}(\nu) \right\|_2 \xrightarrow[M\to\infty]{a.s.} 0
    \end{equation}
Using \eqref{eq:conv_hat_S_y}, we have the bound
\begin{align}
  \max_{\nu \in \Vcal_N} \left\|\dg\left(\hat{\S}_{\y}(\nu)\right) - \dg\left(\S_{\y}(\nu)\right)\right\|_2
  \leq \Delta_1 + \Delta_2,
  \notag
\end{align}
with
\begin{align}
  \Delta_1
  &=
    \max_{\nu \in \Vcal_N} \left\|\hat{\S}_{\y}(\nu) - \S_{\y}(\nu)^{\frac{1}{2}}\frac{\X(\nu)\X(\nu)^*}{B+1}\S_{\y}(\nu)^{\frac{1}{2}}\right\|_2
    \notag\\
  &\xrightarrow[M\to\infty]{a.s.} 0,
    \notag
\end{align}
and
\begin{align}
  &\Delta_2 =
    \max_{\nu \in \Vcal_N}
    \max_{m=1,\ldots,M}
    \left|
    \left[
    \S_{\y}(\nu)^{\frac{1}{2}}\left(\frac{\X(\nu)\X(\nu)^*}{B+1}-\I_M\right)\S_{\y}(\nu)^{\frac{1}{2}}
    \right]_{m,m}
    \right|.
    \notag
\end{align}
Denoting $\u_m(\nu) = \S_{\y}(\nu)^{\frac{1}{2}} \e_m$, where $\e_m$ is the $m$-th vector of the canonical basis of $\Cbb^M$,
as well as $\x_1(\nu),\ldots,\x_{B+1}(\nu)$ the i.i.d. $\Ncal_{\Cbb^M}(\mathbf{0},\I_M)$ column vectors of $\X(\nu)$, we have for all $t > 0$,
\begin{align}
  &\Pbb\left(\Delta_2 > t\right)
  \leq \sum_{\nu \in \Vcal_N} \sum_{m=1}^{M}
  \Pbb\left(\left|\frac{1}{B+1} \sum_{b=1}^{B+1} \left|\u_m(\nu)^*\x_b(\nu)\right|^2 - \|\u_m(\nu)\|_2^2\right| > t\right).
  \notag
\end{align}
From Assumption \ref{assumption:rm}, Assumption \ref{assumption:sm} and condition \eqref{assumption:memory_H} from Assumption \ref{assumption:H}, we have
\begin{align*}
  0
  <
  \inf_{M \geq 1} \min_{m=1,\ldots,M}\min_{\nu \in [0,1]} \left\|\u_m(\nu)\right\|_2 \leq
  \sup_{M \geq 1} \max_{m=1,\ldots,M} \max_{\nu \in [0,1]} \left\|\u_m(\nu)\right\|_2
  < \infty.
  \notag
\end{align*}
Setting in the statement of Lemma \ref{lemma:hanson-wright}
\begin{align}
  \z = \left(\x_1(\nu)^T,\ldots,\x_{B+1}(\nu)^T\right)^T \sim \Ncal_{\Cbb^{M(B+1)}}(\mathbf{0},\I_{M(B+1)})
  \notag
\end{align}
and $\Xibs$ as the $M(B+1)\times M(B+1)$ block-diagonal matrix
\begin{align}
  \Xibs = \frac{\I_{B+1} \otimes \left(\u_m(\nu)\u_m(\nu)^*\right)}{B+1}
  \notag
\end{align}
with $\otimes$ denoting the Kronecker product, we obtain
\begin{align}
  &\Pbb\left(\Delta_2 > t\right)
  \leq  2 M N \max_{\nu \in \Vcal_N}
  \exp\left(-C \min\left\{\frac{B t^2}{\|\u_m(\nu)\|_2^4}, \frac{B t}{\|\u_m(\nu)\|_2^2}\right\}\right)
  \notag
\end{align}
where $C > 0$ is a constant independent of $M$, which in turn implies that
\begin{align}
  \Delta_2 \xrightarrow[M\to\infty]{a.s.} 0
  \notag
\end{align}
and that (\ref{eq:inter-step1}) holds.
In order to check (\ref{eq:inter-step2}), we use Assumption \ref{assumption:H} eq. \eqref{assumption:power_H} to get that
\begin{align}
  \max_{\nu \in \Vcal_N} \left\|\dg\left(\H(\nu)\H(\nu)^*\right)\right\|_2
  &= \max_{\nu \in \Vcal_N}\max_{m=1,\ldots,M} \left\|\h_m(\nu)\right\|_2^2\xrightarrow[M\to\infty]{} 0
    \notag
\end{align}
and from the fact that
\begin{align}
  \dg\left(\S_{\y}(\nu)\right) = \dg\left(\H(\nu)\H(\nu)^*\right) + \S_{\v}(\nu)
  \notag
\end{align}
we obtain \eqref{eq:inter-step2} and, in turn, \eqref{eq:diag_S_hat_y}.

To prove \eqref{eq:diag_S_hat_y_sqrt}, we write (using that $|\sqrt{a}-\sqrt{b}|<\sqrt{|a-b|}$ for $a,b>0$) 
\begin{align*}
  \max_{\nu \in \Vcal_N} \left\|\dg\left(\hat{\S}_{\y}(\nu)\right)^{-\frac{1}{2}} - \S_{\v}(\nu)^{-\frac{1}{2}}\right\|_2 
   \leq
    \max_{\nu \in \Vcal_N} \max_{m=1,\ldots,M}
    \sqrt{\frac{|[\hat{\S}_{\y}(\nu)]_{m,m} - s_m(\nu)|}{[\hat{\S}_{\y}(\nu)]_{m,m}\ s_m(\nu)}}.
\end{align*}
From Assumption \ref{assumption:sm}, there exists $\epsilon > 0$ such that
\begin{align}
  \inf_{M \geq 1} \min_{m=1,\ldots,M}\min_{\nu \in \Vcal_N} s_m(\nu) \geq \epsilon > 0.
  \notag
\end{align}
Using \eqref{eq:diag_S_hat_y} and denoting
\begin{align}
  \Delta = \max_{\nu \in \Vcal_N}\left\|\dg\left(\hat{\S}_{\y}(\nu)\right) - \S_{\v}(\nu)\right\|_2
  \notag
\end{align}
we have that
\begin{align}
  \max_{\nu \in \Vcal_N} \left\|\dg\left(\hat{\S}_{\y}(\nu)\right)^{-\frac{1}{2}} - \S_{\v}(\nu)^{-\frac{1}{2}}\right\|_2
  & \leq
    \sqrt
    {
    \frac
    {
    \Delta
    }
    {
    \epsilon\left(\epsilon - \Delta\right)
    }
    }
    \notag
\end{align}
with probability one for all large $M$, which proves \eqref{eq:diag_S_hat_y_sqrt}.  
\end{IEEEproof}
We also need the following lemma on the boundedness of matrix $\hat{\S}_{\y}(\nu)$.
\begin{lemma}
  \label{lemma:boundedness_S_hat_y}
  Under Assumptions \ref{assumption:rm}, \ref{assumption:H} and \ref{assumption:regime}, we have
  \begin{align}
    \limsup_{M\to\infty} \max_{\nu \in \Vcal_M} \left\|\hat{\S}_{\y}(\nu)\right\|_2 < \infty
    \notag
  \end{align}
  with probability one.
\end{lemma}
\begin{IEEEproof}
  From \eqref{eq:conv_hat_S_y}, we have
  \begin{align}
    &\limsup_{M\to\infty} \max_{\nu \in  \Vcal_N}\left\|\hat{\S}_{\y}(\nu)\right\|_2  \leq
    \limsup_{M\to\infty}
    \max_{\nu \in  \Vcal_N} \left\|\S_{\y}(\nu)\right\|_2
    \max_{\nu \in  \Vcal_N}\left\|\frac{\X(\nu)\X(\nu)^*}{B+1}\right\|_2.
    \notag
  \end{align}
  From Assumptions \ref{assumption:rm} and \ref{assumption:H}, it is clear that
  \begin{align}
    \sup_{M \geq 1} \max_{\nu \in \Vcal_N} \left\|\S_{\y}(\nu)\right\|_2 < \infty.
    \notag
  \end{align}
  Finally, from Lemma \ref{lemma:haagerup} and the remarks below this lemma, we have
  \begin{align}
    \limsup_{M\to\infty}\max_{\nu \in \Vcal_N} \left\|\frac{\X(\nu)\X(\nu)^*}{B+1}\right\|_2 < \infty
    \notag
  \end{align}
  with probability one, and Lemma \ref{lemma:boundedness_S_hat_y} is proved.
\end{IEEEproof}
Equipped with Lemmas \ref{lemma:diag_S_hat_y} and \ref{lemma:boundedness_S_hat_y}, we are now in position to prove Theorem \ref{theorem:C_hat_y}.
Define
\begin{align}
  \tilde{\Delta} = \max_{\nu \in \Vcal_N} \left\|\dg\left(\hat{\S}_{\y}(\nu)\right)^{-\frac{1}{2}} - \S_{\v}(\nu)^{-\frac{1}{2}}\right\|_2
  \notag
\end{align}
and recall the definition of the random matrix $\X(\nu)$ in \eqref{eq:conv_hat_S_y}.
Let us write
\begin{align}
  \hat{\C}_{\y}(\nu) - \Xibs(\nu)^{\frac{1}{2}} \frac{\X(\nu)\X(\nu)^*}{B+1} \Xibs(\nu)^{\frac{1}{2}}
  = \Psi_1(\nu) + \Psi_2(\nu) 
    \notag
\end{align}
where the two error terms are defined by:
\begin{align*}
  \Psi_1(\nu) &= \hat{\C}_\y(\nu) - \S_\v(\nu)^{-\frac{1}{2}}\hat{\S}_\y(\nu)\S_\v(\nu)^{-\frac{1}{2}}
\end{align*}
\begin{align*}
    \Psi_2(\nu) = \S_\v(\nu)^{-\frac{1}{2}}\hat{\S}_\y(\nu)\S_\v(\nu)^{-\frac{1}{2}} - \Xibs(\nu)^{\frac{1}{2}}\frac{\X(\nu)\X^*(\nu)}{B+1}\Xibs(\nu)^{\frac{1}{2}}
\end{align*}
which satisfies:
\[
    \max_{\nu\in\Vcal_N}\left\|\Psi_1(\nu)\right\|_2 \le \tilde{\Delta} \max_{\nu \in \Vcal_N}
  \left\|\hat{\S}_{\y}(\nu)\right\|_2
  \left(
  \tilde{\Delta} + \frac{2}{\sqrt{\min_{\nu \in \Vcal_N} \lambda_M\left(\S_{\v}(\nu)\right)}}
  \right)
    \notag
\]
and
\[
    \max_{\nu\in\Vcal_N}\left\|\Psi_2(\nu)\right\|_2 \le \frac{\max_{\nu \in \Vcal_N}\left\| \hat{\S}_{\y}(\nu) - \S_{\y}(\nu)^{\frac{1}{2}} \frac{\X(\nu)\X(\nu)^*}{B+1} \S_{\y}(\nu)^{\frac{1}{2}} \right\|_2}{\min_{\nu\in\Vcal_N}\lambda_M(\S_\v(\nu))}.
\]

From Assumption \ref{assumption:sm}, we have
\begin{align}
  \inf_{M \geq 1} \min_{\nu \in \Vcal_N} \lambda_M\left(\S_{\v}(\nu)\right) > 0.
  \notag
\end{align}
Using Lemmas \ref{lemma:diag_S_hat_y} and \ref{lemma:boundedness_S_hat_y}, we directly deduce that
\begin{align}
  \max_{\nu\in\Vcal_N}\left\|\Psi_1(\nu)\right\|_2  \xrightarrow[M\to\infty]{a.s.} 0.
  \notag
\end{align}
Likewise, using \eqref{eq:conv_hat_S_y}, we deduce that
\begin{align}
  \max_{\nu\in\Vcal_N}\left\|\Psi_2(\nu)\right\|_2  \xrightarrow[M\to\infty]{a.s.} 0,
  \notag
\end{align}
which concludes the proof of Theorem \ref{theorem:C_hat_y}.

\appendices

\setcounter{section}{3}

\section{Proof of Corollary \ref{corollary:pure_noise}, Corollary \ref{corollary:pure_signal} and Proposition \ref{proposition:S_hat_y}}

\subsection{Proof of Corollary \ref{corollary:pure_noise}}
\label{appendix:proof_corollary_pure_noise}

Write $\hat{\S}_{\v}(\nu) = \Sigmabs_{\v}(\nu)\Sigmabs_{\v}(\nu)^*$, and denote
\begin{align}
  \Delta_{\v}(\nu) = \left\|\Sigmabs_{\v}(\nu) - \frac{1}{\sqrt{B+1}} \S_{\v}(\nu)^{1/2} \Z(\nu)\right\|_2.
  \notag
\end{align}
Using the fact that for any two matrices $\A, \B$ of appropriate dimensions, we have 
\[
    \A\A^* - \B\B^* = (\A-\B)(\A-\B)^* + (\A-\B)\B^* + \B(\A-\B)^*
\]
and
\[
    \|\A\B\|_2 \le \|\A\|_2\|\B\|_2
\]
we see that
\begin{align}
  &\left\|\hat{\S}_{\v}(\nu) - \frac{1}{B+1} \S_{\v}(\nu)^{1/2} \Z(\nu)\Z(\nu)^*\S_{\v}(\nu)^{1/2}\right\|_2 \leq
    \Delta_{\v}(\nu)
    \left(
    \Delta_{\v}(\nu)
    +
    2 \sqrt{\frac{\|\S_{\v}(\nu)\|_2}{B+1}} \|\Z(\nu)\|_2
    \right).
    \notag
\end{align}
Assumption \ref{assumption:rm} implies that
\begin{align}
  \sup_{M \geq 1} \max_{\nu \in [0,1]} \|\S_{\v}(\nu)\|_2 < \infty
  \notag
\end{align}
while from Lemma \ref{lemma:haagerup} from Appendix \ref{appendix:useful_results}, since $\Z(\nu)$ has i.i.d. complex Gaussian entries,
\begin{align}
  \limsup_{M \to \infty} \max_{\nu \in \Vcal_N} \frac{\left\|\Z(\nu)\right\|_2}{\sqrt{B+1}} < \infty
  \label{eq:control_Z}
\end{align}
with probability one. This concludes the proof of Corollary \ref{corollary:pure_noise}.

\subsection{Proof of Corollary \ref{corollary:pure_signal}}
\label{appendix:proof_corollary_pure_signal}

The proof of Corollary \ref{corollary:pure_signal} is similar to the one of Corollary \ref{corollary:pure_noise}.
Denoting $\Delta_{\u}(\nu) = \left\|\Sigmabs_{\u}(\nu) - \H(\nu)\Sigmabs_{\epsilonbs}(\nu)\right\|_2$, and noticing that
$\sup_{M \geq 1} \max_{\nu \in [0,1]} \left\|\H(\nu)\right\|_2 < \infty$ from Assumption \ref{assumption:H} eq. \eqref{assumption:memory_H},
we obtain that
\begin{align}
  &\max_{\nu \in \Vcal_N}\left\|\hat{\S}_{\u}(\nu) - \H(\nu)\Sigmabs_{\epsilonbs}(\nu)\Sigmabs_{\epsilonbs}(\nu)^*\H(\nu)^*\right\|_2
   \leq
    \max_{\nu \in \Vcal_N} \Delta_{\u}(\nu)\left(\Delta_{\u}(\nu) + 2 \left\|\H(\nu)\right\|_2\left\|\Sigmabs_{\epsilonbs}(\nu)\right\|_2\right),
   \xrightarrow[M\to\infty]{a.s.} 0.
    \notag
\end{align}
Since $K$ is fixed with respect to $M$ from Assumption \ref{assumption:regime}, we also have
\begin{align}
  \max_{\nu \in \Vcal_N} \left\|\Sigmabs_{\epsilonbs}(\nu)\Sigmabs_{\epsilonbs}(\nu)^* - \I_M\right\|_2
  \xrightarrow[M\to\infty]{a.s.} 0
  \label{eq:control_sigma_epsilon}
\end{align}
using Lemma \ref{lemma:hanson-wright}, which proves Corollary \ref{corollary:pure_signal}.

\subsection{Proof of Proposition \ref{proposition:S_hat_y}}
\label{appendix:proof_proposition_S_hat_y}

To prove Proposition \ref{proposition:S_hat_y}, let us write
\begin{align}
  \Y(\nu) = \H(\nu)\Sigmabs_{\epsilonbs}(\nu) + \frac{\S_{\v}(\nu)^{1/2} \Z(\nu)}{\sqrt{B+1}}.
  \notag
\end{align}
Then, from \eqref{eq:control_Z} and \eqref{eq:control_sigma_epsilon}, we have
\begin{align}
  \limsup_{M \to \infty} \max_{\nu \in \Vcal_N} \left\|\Y(\nu)\right\|_2 < \infty
  \notag
\end{align}
with probability one, which implies the following convergence
\begin{align}
  &\max_{\nu \in \Vcal_N}\left\|\hat{\S}_{\y}(\nu) - \Y(\nu)\Y(\nu)^*\right\|_2
  \leq
    \max_{\nu \in \Vcal_N} \left(\Delta_{\u}(\nu) + \Delta_{\v}(\nu)\right) \left(\Delta_{\u}(\nu) + \Delta_{\v}(\nu) + 2 \left\|\Y(\nu)\right\|_2\right)
    \xrightarrow[M\to\infty]{a.s.} 0.
    \notag
\end{align}
Finally, since the columns of $\sqrt{B+1}\Y(\nu)$ are i.i.d. $\Ncal_{\Cbb^M}(\mathbf{0},\S_{\y}(\nu))$ with $\S_{\y}(\nu) = \H(\nu)\H(\nu)^* + \S_{\v}(\nu)$, it follows that
\begin{align}
  \Y(\nu) = \S_{\y}(\nu)^{1/2} \frac{\X(\nu)}{\sqrt{B+1}}
  \notag
\end{align}
for some $M \times (B+1)$ matrix $\X(\nu)$ having i.i.d. $\Ncal_{\Cbb}(0,1)$ entries and the proof of Proposition \ref{proposition:S_hat_y} is complete.

\section{Proof of Corollary \ref{corollary:lss}}
\label{appendix:proof_corollary_lss}

We first prove that all the eigenvalues of the SCM asymptotically concentrate in a compact set with probability one for all large $M$.
Indeed, considering matrix
$$\W(\nu) = \Xibs(\nu)^{\frac{1}{2}} \frac{\X(\nu)\X(\nu)^*}{B+1}\Xibs(\nu)^{\frac{1}{2}}$$
defined through Theorem \ref{theorem:C_hat_y} and using Lemma \ref{lemma:haagerup} in conjunction with Borel-Cantelli lemma, we deduce that there exists
constants $C_1,C_2$ such that
\begin{align}
  \liminf_{M\to\infty} \min_{\nu \in \Vcal_N}\lambda_M\left(\W(\nu)\right) \geq C_1 (1-\sqrt{c})^2
  \notag
\end{align}
and
\begin{align}
  \limsup_{M\to\infty} \max_{\nu \in \Vcal_N}\lambda_1\left(\W(\nu)\right)\leq C_2 (1+\sqrt{c})^2
  \label{eq:upper_bound_W}
\end{align}
with probability one, where $C_1,C_2$ verify, thanks to Assumption \ref{assumption:H},
\begin{align}
  0 < C_1 < 1 = \inf_{M \geq 1} \min_{\nu \in \Vcal_N} \lambda_M\left(\Xibs(\nu)\right)
  \notag
\end{align}
and
\begin{align}
  \sup_{M \geq 1} \max_{\nu \in \Vcal_N} \lambda_M\left(\Xibs(\nu)\right) < C_2 < \infty.
  \notag
\end{align}
Using \eqref{eq:conv_eig}, we obtain similarly
$$\liminf_{M\to\infty} \min_{\nu \in \Vcal_N}\lambda_M\left(\hat{\C}_{\y}(\nu)\right) \geq C_1 (1-\sqrt{c})^2$$
and
$$\limsup_{M\to\infty} \max_{\nu \in \Vcal_N}\lambda_1\left(\hat{\C}_{\y}(\nu)\right)\leq C_2 (1+\sqrt{c})^2$$
with probability one.
Let $0 < \epsilon < \frac{C_1}{2} (1-\sqrt{c})^2$ and $h \in \Ccal_c^{1}(\Rbb)$ such that
\begin{align}
  h(\lambda) =
  \begin{cases}
    1 & \text{ if } \lambda \in \left[C_1 (1-\sqrt{c})^2 - \epsilon, C_2 (1+\sqrt{c})^2 + \epsilon\right]
    \notag\\
    0 & \text{ if } \lambda \not\in \left[C_1 (1-\sqrt{c})^2 - 2\epsilon, C_2 (1+\sqrt{c})^2 + 2\epsilon\right]
  \end{cases}.
\end{align}
Then it follows that
\begin{align}
  \max_{\nu \in \Vcal_N} \left|L_{\varphi}(\nu) - L_{\varphi h}(\nu)\right| \xrightarrow[M\to\infty]{a.s.} 0.
  \notag
\end{align}
Thus, without loss of generality, we may assume for the remainder of the proof that $\varphi \in \Ccal^1_c\left((0,+\infty)\right)$.
Using \eqref{eq:conv_eig}, we deduce that
\begin{align}
  \max_{\nu \in \Vcal_N} \frac{1}{M}\sum_{m=1}^M \left|\varphi\left(\lambda_m\left(\hat{\C}_{\y}(\nu)\right)\right) - \varphi\left(\lambda_m\left(\W(\nu)\right)\right)\right|
  \xrightarrow[M\to\infty]{a.s.} 0.
  \notag
\end{align}
Next, consider the two functions
\begin{align}
  \hat{m}(z,\nu)
  =
  \frac{1}{M} \sum_{m=1}^M \frac{1}{\lambda_m\left(\W(\nu)\right) - z}
  = \int_{\Rbb} \frac{\drm \hat{\mu}(\lambda,\nu)}{\lambda - z}
  \notag
\end{align}
and
\begin{align}
  \tilde{m}(z,\nu)
  = \frac{1}{M} \sum_{m=1}^M \frac{1}{\lambda_m\left(\frac{\X(\nu)\X(\nu)^*}{B+1}\right) - z}
  = \int_{\Rbb} \frac{\drm \tilde{\mu}(\lambda,\nu)}{\lambda - z}
  \notag
\end{align}
defined for all $z \in \Cbb^+ := \{\zeta \in \Cbb: \Im(\zeta) > 0\}$, and where for all Borel set $A \subset \Rbb$,
\begin{align}
  \hat{\mu}(A,\nu) = \frac{1}{M} \sum_{m=1}^M \delta_{\lambda_m\left(\W(\nu)\right)}(A)
  \notag
\end{align}
and
\begin{align}
  \tilde{\mu}(A,\nu) = \frac{1}{M} \sum_{m=1}^M \delta_{\lambda_m\left(\frac{\X(\nu)\X(\nu)^*}{B+1}\right)}(A)
  \notag
\end{align}
denote the empirical eigenvalue distributions of matrices $\W(\nu)$ and $\frac{\X(\nu)\X(\nu)^*}{B+1}$ respectively, and $\delta_x$ is the Dirac measure at point $x$.
Functions $z \mapsto \hat{m}(z,\nu)$ and $z \mapsto \tilde{m}(z,\nu)$ coincide with the Stieltjes transforms of measures $\hat{\mu}(.,\nu)$ and $\tilde{\mu}(.,\nu)$
respectively (see \cite{Tao2012} for a review of the main properties of the Stieltjes transform).
Since
\begin{align}
  & \hat{m}(z,\nu) - \tilde{m}(z,\nu) = 
    \frac{1}{M} \tr \left(\left(\W(\nu)-z\I\right)^{-1} - \left(\frac{\X(\nu)\X(\nu)^*}{B+1}-z\I\right)^{-1}\right)
     \notag
\end{align}
and using the fact that $\A^{-1}-\B^{-1} = \A^{-1}(\B-\A)\B^{-1}$ for non-singular matrices $\A,\B$, we have
\begin{align}
  &\left|\hat{m}(z,\nu) - \tilde{m}(z,\nu)\right| \leq
\frac{1}{|\Im(z)|^2}
  \frac{K}{M} \left\|\H(\nu)\right\|_2^2 \left\|\S_{\v}(\nu)^{-1}\right\|_2
  \left\|
  \frac{\X(\nu)\X(\nu)^*}{B+1}
  \right\|_2
    \notag
\end{align}
it follows from Assumptions \ref{assumption:sm}, \ref{assumption:H}, \ref{assumption:regime} and Lemma \ref{lemma:haagerup} that
\begin{align}
  \max_{\nu \in \Vcal_N} \left|\hat{m}(z,\nu) - \tilde{m}(z,\nu)\right| \xrightarrow[M\to\infty]{a.s.} 0
  \label{eq:conv_st}
\end{align}
for all $z\in \Cbb^+$.
In the following, we fix a realization in an event of probability one for which \eqref{eq:conv_st} holds for all $z \in \Cbb^+$ and consider
\begin{align}
  \nu^* \in \argmax_{\nu \in \Vcal_N} \left|\int_{\Rbb} \varphi(\lambda) \drm \hat{\mu}(\lambda,\nu) - \int_{\Rbb} \varphi(\lambda) \drm \tilde{\mu}(\lambda,\nu)\right|.
  \notag
\end{align}
Then $\left|\hat{m}(z,\nu^*) - \tilde{m}(z,\nu^*)\right| \to  0$ as $M\to\infty$, for all $z \in  \Cbb^+$.
From the fact that the pointwise convergence on $\Cbb^+$ of a sequence of Stieltjes transforms is equivalent to the weak convergence of the related sequence of probability measures (see e.g. \cite[Ex.2.4.10]{Tao2012}), we deduce that
\begin{align}
  &\max_{\nu \in \Vcal_N}
  \left|\frac{1}{M} \sum_{m=1}^M \left(\varphi\left(\lambda_m\left(\W(\nu)\right)\right)
    - \varphi\left(\lambda_m\left(\frac{\X(\nu)\X(\nu)^*}{B+1}\right)\right)\right)\right|
    = \left|\int_{\Rbb} \varphi(\lambda) \drm \hat{\mu}(\lambda,\nu^*) - \int_{\Rbb} \varphi(\lambda) \drm \tilde{\mu}(\lambda,\nu^*)\right|
    \xrightarrow[M\to\infty]{} 0.
  \notag
\end{align}
To conclude the proof of Corollary \ref{corollary:lss}, it remains to prove that
\begin{align}
  &\max_{\nu \in \Vcal_N}
  \left|\frac{1}{M} \sum_{m=1}^M \varphi\left(\lambda_m\left(\frac{\X(\nu)\X(\nu)^*}{B+1}\right)\right) - \int_{\Rbb} \varphi(\lambda) f(\lambda) \drm\lambda\right|
    \xrightarrow[M\to\infty]{a.s.} 0.
  \notag
\end{align}
Consider the decomposition
\begin{align}
  \max_{\nu \in \Vcal_N}
  \left|\frac{1}{M} \sum_{m=1}^M \varphi\left(\lambda_m\left(\frac{\X(\nu)\X(\nu)^*}{B+1}\right)\right)
  - \int_{\Rbb} \varphi(\lambda) f(\lambda) \drm\lambda\right| \leq \Delta_1 + \Delta_2,
  \notag
\end{align}
where
\begin{align*}
  \Delta_1 =
  \max_{\nu \in \Vcal_N}
  \Biggl|
  \frac{1}{M} \sum_{m=1}^M \Biggl(\varphi\left(\lambda_m\left(\frac{\X(\nu)\X(\nu)^*}{B+1}\right)\right)
  - \Ebb\left[\varphi\left(\lambda_m\left(\frac{\X(\nu)\X(\nu)^*}{B+1}\right)\right)\right]\Biggr)
  \Biggr|
\end{align*}
and
\begin{align}
\notag
  \Delta_2 =  \left|\frac{1}{M} \sum_{m=1}^M \Ebb\left[\varphi\left(\lambda_m\left(\frac{\X(0)\X(0)^*}{B+1}\right)\right)\right] - \int_{\Rbb} \varphi(\lambda) f(\lambda) \drm\lambda\right|.
  \notag
\end{align}
Using the concentration inequality of \cite[Cor. 1.8(b)]{Guionnet2000}, it is straightforward to show that
\begin{align}
  \Delta_1 \xrightarrow[M\to\infty]{a.s.} 0.
  \notag
\end{align}
Moreover, using again the properties of the Stieltjes transform, it can be deduced from
e.g. \cite{Hachem2008} that
\begin{align}
  \Delta_2 \xrightarrow[M\to\infty]{} 0.
  \notag
\end{align}
This concludes the proof of Corollary \ref{corollary:lss}.

\section{Proof of Proposition \ref{proposition:spike}}
\label{appendix:proof_proposition_spike}

Convergences \eqref{eq:spikelimit1}, \eqref{eq:spikelimit2} and \eqref{eq:spikelimit3} are straightforward consequences of \eqref{eq:conv_eig}
and the results of \cite[Th. 1.1]{Baik2006} on the behaviour of the largest eigenvalues for the so-called multiplicative spike model random matrices.
To prove \eqref{eq:spikelimit4}, we use the bound
\begin{align}
  \lambda_1\left(\W(\nu)\right) \leq \lambda_1\left(\frac{\X(\nu)\X(\nu)^*}{B+1}\right) \lambda_1\left(\Xibs(\nu)\right)
  \notag
\end{align}
Then, from the fact that $\gamma_{\infty} = 0$ and Lemma \ref{lemma:haagerup}, we finally obtain
\begin{align}
  \limsup_{M\to\infty} \max_{\nu \in \Vcal_N} \lambda_1\left(\W(\nu)\right)
  &\leq \limsup_{M\to\infty} \max_{\nu \in \Vcal_N} \lambda_1\left(\frac{\X(\nu)\X(\nu)^*}{B+1}\right)
    \notag\\
  & \leq \left(1+\sqrt{c}\right)^2.
    \notag
\end{align}
The proof is concluded by invoking again convergence \eqref{eq:conv_eig}.

\end{appendices}

\end{document}